\documentclass[prb,amsmath,showpacs,twocolumn,floatfix]{revtex4-1}

\usepackage{float}
\usepackage{amssymb}
\usepackage{graphicx}
\usepackage{subfigure}
\usepackage{color}
\usepackage{multirow}
\usepackage{verbatim}

\usepackage{dcolumn}

\begin{document}

\title{First-principles model potentials for lattice-dynamical
  studies: general methodology and example of application to ferroic
  perovskite oxides}

\author{Jacek C. Wojde\l,$^{1}$ Patrick Hermet,$^{2,3}$ Mathias
  P. Ljungberg,$^{1}$ Philippe Ghosez,$^{2}$ and Jorge
  \'I\~niguez$^{1}$}

\affiliation{$^{1}$Institut de Ci\`encia de Materials de Barcelona
  (ICMAB-CSIC), Campus UAB, 08193 Bellaterra, Spain\\$^{2}$Physique
  Th\'eorique des Mat\'eriaux, Universit\'e de Li\`ege (B5), B-4000
  Li\`ege, Belgium\\$^{3}$Institut Charles Gerhardt Montpellier, UMR
  5253 CNRS-UM2, Equipe C$_{2}$M, Universit\'e Montpellier 2, Place
  Eug\`ene Bataillon, cc1504, 34095 Montpellier Cedex 5, France}

\begin{abstract}
We present a scheme to construct model potentials, with parameters
computed from first principles, for large-scale lattice-dynamical
simulations of materials. Our method mimics the traditional
solid-state approach to the investigation of vibrational spectra,
i.e., we start from a suitably chosen reference configuration of the
material and describe its energy as a function of arbitrary atomic
distortions by means of a Taylor series. Such a form of the
potential-energy surface is completely general, trivial to formulate
for any compound, and physically transparent. Further, the
approximations involved in our effective models -- i.e., the
truncations affecting the order of the polynomial expansion, the
spatial range of the interatomic couplings, and the maximum number of
atoms (or {\em bodies}) involved in the interaction terms of the
series -- are clear-cut, and the precision can be improved in a
systematic and well-defined fashion. Moreover, such a simple
definition allows for a straightforward determination of the
parameters in the low-order terms of the series, as they are the
direct result of density-functional-perturbation-theory calculations,
which greatly simplifies the model construction. Here we present such
a scheme, discuss a practical and versatile methodology for the
calculation of the model parameters from first principles, and
describe our results for two challenging cases in which the model
potential is strongly anharmonic, namely, ferroic perovskite oxides
PbTiO$_{3}$ and SrTiO$_{3}$. The choice of test materials was partly
motivated by historical reasons, since our scheme can be viewed as a
natural extension of (and was initially inspired by) the so-called
first-principles {\em effective Hamiltonian} approach to the
investigation of temperature-driven effects in ferroelectric
perovskite oxides. Thus, the study of these compounds allows us to
better describe the connections between the effective-Hamiltonian
method and ours.
\end{abstract}






\maketitle

\section{Introduction}

The development of methods for statistical simulations with
first-principles accuracy remains one of the major challenges for the
community working on computational condensed-matter physics and
materials science. In spite of recent advances, state-of-the-art
first-principles methods are still unable to reach the length and time
scales that are relevant for the study of many properties of interest
at realistic operating conditions. Ranging from temperature-driven
phase transitions to thermally-activated processes of all sorts, there
are countless phenomena whose first-principles treatment has a
prohibitive computational cost, even if one resorts to the most
numerically-efficient schemes such as density-functional theory
(DFT).\cite{martin-book2004} Hence, there is a need to develop
approximate methods that allow for fast calculations while retaining
the first-principles accuracy and, if possible, predictive power. Much
of the on-going activity on multi-scale simulations is the direct
consequence of this situation.

Whenever one is concerned with the lattice-dynamical properties of the
materials, it may be possible to avoid the explicit treatment of the
electrons in the simulations. Such is typically the case when we are
interested in structural and mechanical properties, dielectric and
piezoelectric responses (which are dominated by the lattice part of
the effect, as opposed to the electronic one, in the materials that
are most attractive for applications), or lattice thermal transport,
to name a few important examples. Many methods have been developed to
address this subset of problems, which are the focus of the present
work.

In the context of lattice-dynamical studies, there are essentially two
families of effective potentials that allow for large-scale
simulations. The most widely used methods have been developed in the
Physics and Chemistry communities. Such models -- which include
Lennard-Jones potentials,\cite{ashcroft-book1976} shell
models,\cite{sepliarsky05} bond-valence models,\cite{brown09,shin05}
and even Tersoff potentials\cite{tersoff88} and reactive force
fields,\cite{vanduin01} to name a few -- tend to rely on a
physically-motivated analytic form of the atomic
interactions. Unfortunately, that restriction is often a too stringent
one, and compromises the ability of the models to reproduce the
first-principles data. Further, a systematic extension of the models
to improve precision is usually not well defined or possible.

Another approach is represented by the methods that rely on artificial
neural networks,\cite{jovanjose12} importing techniques developed by
the artificial intelligence community. In this case, one uses very
versatile models that can reproduce first-principles data with
arbitrary precision, at the expense of creating complicated potentials
that do not allow for a clear physical interpretation. On top of the
loss in fundamental understanding, the fact that such models are not
physically motivated usually implies that they have poor
transferability and a limited predictive power [i.e., they are good
  for interpolating between the first-principles data points used to
  fit the model, but often fail when used to predict (extrapolate) new
  behaviors]. Additionally, they are relatively costly from a
computational point of view, as the potentials can become quite
complex.

Interestingly, some workers have developed alternative, very
successful approaches that overcome most of the above mentioned
deficiencies, but which have been applied to a very small set of
problems. One relevant example is the work of Rabe, Vanderbilt, and
others on ferroelectric perovskite oxides: Already in the 1990s these
authors constructed first-principles model potentials, which are
usually called {\em effective Hamiltonians}, to investigate the
ferroelectric phase transitions of perovskites like
BaTiO$_{3}$\cite{zhong94a,zhong95a} and PbTiO$_{3}$.\cite{waghmare97}
[These works built upon the ideas introduced in
  Ref.~\onlinecite{rabe87} to investigate the structural phase
  transition in GeTe from first principles.] The effective-Hamiltonian
approach involves a drastic simplification of the material, which is
coarse-grained to retain only those degrees of freedom associated with
the ferroelectric properties (i.e., local dipoles and cell
strains). The potential-energy surface (PES) corresponding to these
relevant variables is written as a low-order Taylor series around a
suitably chosen reference structure (i.e., the prototype cubic
perovskite structure). Such a scheme is physically-motivated,
computationally very efficient, and its precision can be improved, to
some extent, in a well-defined way. Further, the application of the
original approach to increasingly complex oxides (e.g., compounds with
non-polar transitions like SrTiO$_3$,\cite{zhong95b,zhong96}
chemically-disordered materials like
PbZr$_{1-x}$Ti$_{x}$O$_{3}$,\cite{bellaiche00,kornev06} and
magnetoelectric multiferroics like
BiFeO$_{3}$\cite{kornev07,prosandeev13}) has shown its generality, the
good transferability of the interatomic couplings among dissimilar
chemical environments, and the reliability and predictive power of the
models. Unfortunately, as far as we know, such an approach has not
been adopted in other research fields, remaining much confined within
a small community working on ferroic perovskite oxides.

In our opinion, to understand why the effective-Hamiltonian method has
failed to gain widespread popularity, one has to consider the
coarse-graining step involved in the construction of the
potential. When these models were first developed, there were plenty
of reasons to adopt such a simplification. On the one hand, by
restricting to a subset of the configuration space, it is possible to
construct simpler potentials and run faster simulations. On the other
hand, by the time first-principles methods started to be applied to
these problems, there was already a whole body of literature devoted
to similar, semi-empirical models used in theoretical studies of phase
transitions driven by soft
modes.\cite{pytte72,blinc-book1974,lines-book1977} Indeed, the
effective Hamiltonians of Refs. \onlinecite{zhong94a} and
\onlinecite{waghmare97} can be viewed as the natural evolution of the
models that already existed in the literature, as for example the
so-called discrete $\phi^{4}$ model.\cite{rubtsov00} In some sense,
the main innovation in those pioneering works was to develop a
systematic and well-defined scheme to compute the parameters of such
Hamiltonians from first principles. To do that, the key step was to
establish a connection between the variables of the traditional
effective models (i.e., the so-called {\em local modes} representing
the localized atomic distortions whose collective occurrence leads to
the structural transition, and which involve the formation of local
electric dipoles in the case of ferroelectrics) and the displacements
of the actual atoms in the crystal. Such a connection can be made in a
variety of ways, ranging from the more elementary\cite{zhong95a} (e.g.,
by defining the local modes from direct inspection of the strongest
structural instabilities of the high-symmetry phase, which can be
determined from first principles as discussed below) to the more
sophisticated\cite{rabe95,iniguez00} (e.g., by identifying the local
modes with {\em lattice} Wannier functions computed from knowledge of
the full phonon dispersion bands of the high-symmetry phase as
obtained from first principles). Once the local modes are defined in
terms of actual atomic displacements, the first-principles calculation
of the Hamiltonian parameters follows in a rather straightforward way.

However, in general there are no clear reasons to introduce such a
coarse graining. Suppose, for example, that we want to investigate a
stable phase of a material, and need a model that captures the
first-principles energetics with very high precision. In such a case,
in absence of structural instabilities of our reference configuration,
it may be unclear how to choose a subset of relevant degrees of
freedom. Further, we may typically find that most of the modes, even
the relatively high-energy ones, play an important role in determining
the properties of such a phase at a {\em quantitative} level; hence,
the restriction to a configuration subspace, and the reduced accuracy
it entails, may not be acceptable. Even in cases where the focus is on
the investigation of structural phase transitions, and assuming that
we would be satisfied by a sound qualitative description of such a
drastic effect, the coarse-graining step may turn out to be both
unhelpful and difficult to implement. Consider, for example, the
modern perovskite oxide super-lattices that present a wealth of
appealing physical effects, some of which are attributed to novel
interatomic couplings occurring at the interfaces between different
layers. In such situations, suitably exemplified by the short-period
PbTiO$_{3}$/SrTiO$_{3}$ super-lattices studied by some of
us,\cite{bousquet08} identifying a small set of relevant degrees of
freedom may be very hard; indeed, a large number of local modes may
need to be considered, which would result in complicated effective
models and a relatively small gain in computational efficiency. Other
cases where similar difficulties are likely to appear are crystals in
which the relative stability of different phases depends strongly on
{\em secondary} structural order parameters (as in the case of
BiFeO$_{3}$\cite{dieguez11}), situations in which strong strain
gradients and non-trivial structural relaxations occur (as in the
vicinity of ferroelectric domain walls\cite{catalan11}), etc. In our
opinion, the local-mode approximation is not well suited for the study
of such problems.

In view of this, we decided to adopt an approach that retains many of
the good features of the effective-Hamiltonian method developed within
the ferroelectrics community {\em and} avoids its most serious
limitations. In short, we decided to create models that describe the
energetics of {\em all} the atomic degrees of freedom of a material by
Taylor expanding the PES around a suitably chosen reference
structure. Working with a simple polynomial model has many important
advantages: ($i$) it is general and can be trivially formulated for
any material; ($ii$) the involved parameters have an obvious and
convenient physical interpretation, as we essentially follow the
approach adopted in solid-state textbooks to discuss lattice-dynamical
and elastic properties; ($iii$) many of the potential parameters can
be obtained directly from perturbative first-principles calculations;
and ($iv$) the precision of the potential can be improved in a
systematic and well-defined way. Here we describe the details of such
a scheme and illustrate it with applications to two challenging cases:
ferroics PbTiO$_3$ and SrTiO$_3$, both of which present
soft-mode-driven structural phase transitions whose description
requires the use of strongly anharmonic potentials. We are thus
introducing a method that we think should be very useful and of
general applicability.

The paper is organized as follows. In Section~II we describe the
general methodology, using perovskite oxides as an illustrative sample
case. We also introduce the approach we adopted to compute the
potential parameters from first principles. In Section~III we describe
the models constructed for perovskite oxides PbTiO$_{3}$ and
SrTiO$_{3}$. We also solve the models by means of Monte Carlo
simulations, showing that they reproduce the experimentally observed
phase transitions. Finally, in Section~IV we summarize and conclude
the paper.

\section{model construction}

In the following we present our general scheme for constructing
effective model potentials for lattice-dynamical studies. The proposed
methodology is general and can be applied to any material, including
cases of reduced dimensionality (e.g., surfaces, slabs, wires,
molecules), disordered systems, etc. Nevertheless, for the sake of
clarity, here we will refer to the case of an infinite periodic crystal, and
take the family of perovskite oxides as a representative example of
application.

\subsection{Reference structure and model variables}

The construction of our models begins with the choice of a reference
structure (RS) that will typically be a minimum or a saddle point of
the PES. Thus, for example, if we were interested in the properties of
a particular (meta)stable phase of a material, the RS would correspond
to the solution obtained for such a phase by performing a
first-principles structural relaxation nominally at $T$~=~0~K. If we
were interested in the more challenging case of a material undergoing
structural phase transitions driven by the {\em softening} (i.e.,
destabilization) of some vibrational modes or cell strains, it would
be convenient to take the high-symmetry phase of the material as our
RS. More specifically, in that case we would determine the RS by
performing a {\em constrained} relaxation (i.e., one in which the high
symmetry of the undistorted phase is imposed) using first-principles
calculations at $T$~=~0~K; the result will typically be a saddle point
of the PES, and will have an associated Hessian matrix (i.e., a matrix
of second derivatives of the energy) with negative eigenvalues
corresponding to the structural instabilities. Our chosen examples of
application -- i.e., ferroics PbTiO$_3$ and SrTiO$_3$ -- belong to
this second category.

For the sake of concreteness, let us assume that we are treating a
three-dimensional infinite crystal composed of periodically repeated
cells. Then, our RS is fully specified by the lattice vectors
$\boldsymbol{R}_{l}$, where $l$ labels cells, and the positions
$\boldsymbol{\tau}_{\kappa}$ of the atoms $\kappa$ inside the cell. We
will describe all accessible crystal configurations as distortions of
the RS. The most general atomic state will be given by the position
vectors
\begin{equation}
 r_{l\kappa\alpha} = \sum_{\beta}
 (\delta_{\alpha\beta}+\eta_{\alpha\beta})(R_{l\beta} +
 \tau_{\kappa\beta})+ u_{l\kappa\alpha } \, ,
\label{eq:atomic-state}
\end{equation}
where $\alpha$ and $\beta$ denote Cartesian directions. The
distortions are thus captured by the homogeneous strain
$\boldsymbol{\eta}$ and the individual atomic displacements
$\boldsymbol{u}_{l\kappa}$. It is important to note that the
homogeneous strain affects both the $\boldsymbol{R}_{l}$ and
$\boldsymbol{\tau}_{\kappa}$ vectors defining the RS, and that we
describe the deviation from the strained RS by means of the {\em
  absolute} displacements $\boldsymbol{u}_{l\kappa}$. Hence, the
$\boldsymbol{u}_{l\kappa}$ vectors are given in Cartesian coordinates
and have units of length. Alternatively, we could have worked with
atomic displacements given in units relative to the (strained)
reference structure; however, our definition of the
$\boldsymbol{u}_{l\kappa}$ variables leads to a clear and
computationally convenient formulation of the model potentials, which
is why we adopted it.

The homogeneous strain $\eta_{\alpha\beta}$ in
Eq.~(\ref{eq:atomic-state}) contains both symmetric and anti-symmetric
parts. Typically, it will be convenient to exclude the anti-symmetric
part (i.e., rigid rotations of the whole material) from the
description. We will thus restrict ourselves to the symmetric
components $(\eta_{\alpha\beta}+\eta_{\beta\alpha})/2$, for which we
will use the well-known Voigt notation $\eta_{a}$, with $a$~=~1, ...,
6.\cite{nye-book1985}

To alleviate the notation in the formulas below, it is convenient to
introduce the following bijective mapping
\begin{equation}
l\kappa \leftrightarrow i \, ,
\label{eq:index-mapping}
\end{equation}
so that any $l\kappa$ pair can be expressed by a single index $i$, and
{\sl vice versa}. Hence, we can use $\boldsymbol{u}_{i}$ instead of
$\boldsymbol{u}_{l\kappa}$, and even write $\boldsymbol{R}_{i}$ or
$\boldsymbol{\tau}_{i}$, without any ambiguity.

\begin{figure}
\includegraphics[width=0.9\columnwidth]{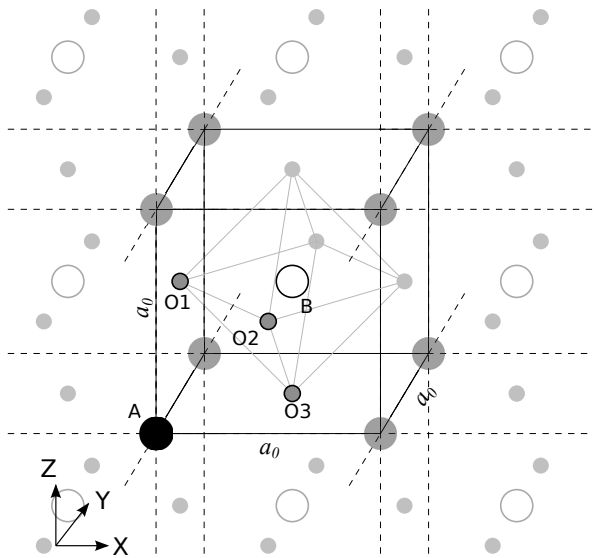}
\caption{Definition of the cubic reference structure for ABO$_{3}$
  perovskite oxides. The unit cell vectors are $\boldsymbol{a}$ =
  $a_{0}(1,0,0)$, $\boldsymbol{b}$ = $a_{0}(0,1,0)$, and
  $\boldsymbol{c}$ = $a_{0}(0,0,1)$, as expressed in the Cartesian
  reference depicted in the figure. Lattice vectors are given by
  $\boldsymbol{R}_{l}$ = $n_{l1}\boldsymbol{a}$ +
  $n_{l2}\boldsymbol{b}$ + $n_{l3}\boldsymbol{c}$, where $n_{l1}$,
  $n_{l2}$, and $n_{l3}$ are integers. The positions of the atoms
  within the unit cell are: $\boldsymbol{\tau}_{\rm A}$ =
  $a_{0}(0,0,0)$, $\boldsymbol{\tau}_{\rm B}$ = $a_{0}(1/2,1/2,1/2)$,
  $\boldsymbol{\tau}_{\rm O1}$ = $a_{0}(0,1/2,1/2)$,
  $\boldsymbol{\tau}_{\rm O2}$ = $a_{0}(1/2,0,1/2)$, and
  $\boldsymbol{\tau}_{\rm O3}$ = $a_{0}(1/2,1/2,0)$.}
\label{fig:model-perovskite}
\end{figure}

Figure~\ref{fig:model-perovskite} shows the cubic phase of an
ABO$_{3}$ perovskite oxide, which is the RS of the applications
discussed below. The shown cell is repeated along the three spatial
directions and, while the displacements $\boldsymbol{u}_{i}$ may
change from cell to cell, the homogeneous strain $\boldsymbol{\eta}$
is the same throughout the material.

\begin{figure*}
\includegraphics[width=\textwidth]{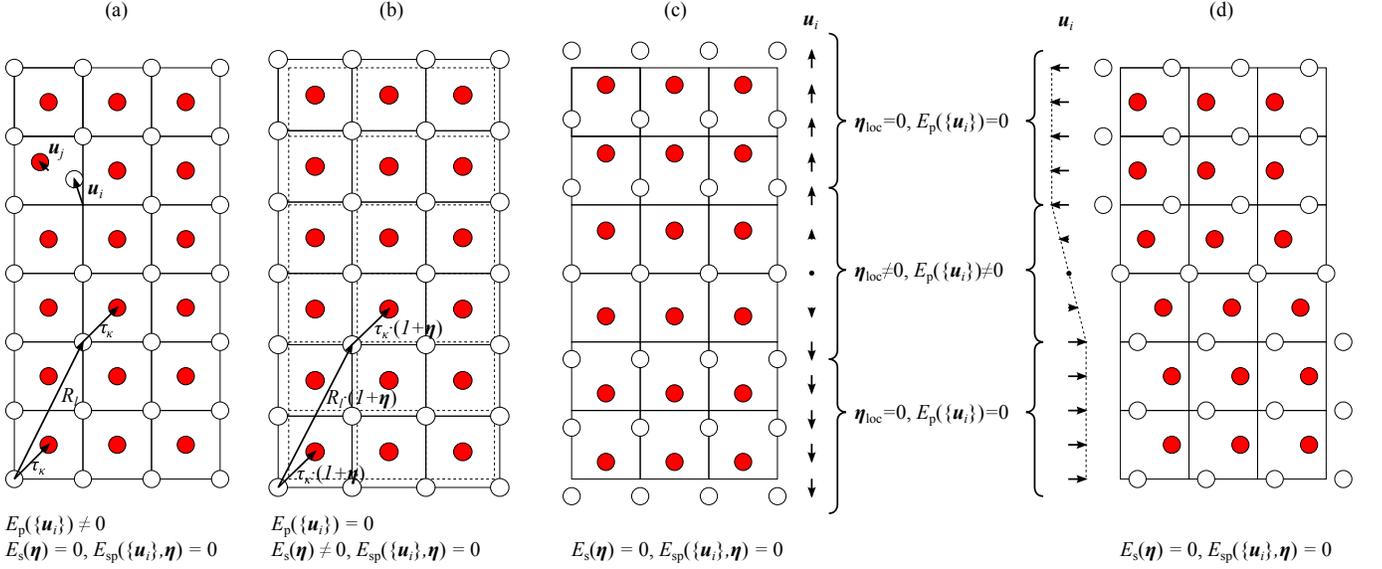}
 \caption{(Color on line.) Sketch of various types of distortions and
   how the associated energy changes are captured by our model
   potentials. Panel~(a): atomic displacements in absence of
   homogeneous strain. Panel~(b): homogeneous strain in absence of
   atomic displacements. Panels~(c) and (d): local (inhomogeneous)
   strain $\boldsymbol{\eta}_{\rm loc}$ given by atomic displacement
   patterns; note that the distortions in the unstrained areas are
   rigid translations, which do not contribute to the energy because
   of the ASR satisfied by $E_{\rm p}(\{\boldsymbol{u}_{i}\})$ (see
   the text).}
\label{fig:strain-decomp}
\end{figure*}

In Fig.~\ref{fig:strain-decomp} we illustrate how an arbitrary
distortion is captured by the variables defined in
Eq.~(\ref{eq:atomic-state}). Panel~(a) shows an unstrained
configuration ($\boldsymbol{\eta}$~=~0) with atoms fluctuating around
their RS positions. Panel~(b) shows an homogeneously strained state
with the atoms maintaining their relative positions
($\boldsymbol{u}_{i}$~=~0). Finally, panels~(c) and (d) show
configurations in which, while the homogeneous strain is zero, we do
have local strains resulting in the expansion along the vertical
direction of some cells of the material [(c)] or a shear-like
deformation [(d)]. Note that such {\em inhomogeneous} strains are
naturally described by the $\boldsymbol{u}_{i}$ variables in our
model.

Before we continue, it is worth mentioning the advantages and
limitations of constructing model potentials based on a reference
structure. As will be obvious below, this approach makes it trivial to
formulate the potential for arbitrary materials and with arbitrary
precision. Hence, it allows for a general and clear-cut methodology
that offers the possibility of improving the models systematically;
those are obviously very important assets that are not so frequently
found in model-potential schemes. At the same time, our approach is
specifically suited for the description of relatively small
distortions of the RS. In other words, we expect our effective models
to describe the energy of configurations that resemble the RS in some
fundamental way; for example, the bond topology should be roughly
preserved, and we would not advice the use of the present method to
describe situations in which chemical bonds break or form.

Finally, let us stress that our scheme is applicable to any material,
not only to the infinite three-dimensional crystals that we focus on
for the sake of the presentation. Indeed, materials of arbitrary
dimensionality, or disordered compounds, can be tackled by making the
appropriate adjustments. For example, to study a molecule we would
work with atomic positions defined by $r_{\kappa\alpha}$ =
$\tau_{\kappa\alpha} + u_{\kappa\alpha }$; to work with materials that
are periodic only along one or two directions, we would just need to
consider an appropriate homogeneous strain tensor; to deal with
chemically-disordered materials, we would construct potentials that
depend on the chemical environment of the atoms, etc. While some
situations may be more challenging than others, our methodology can in
principle be applied in all cases.

\subsection{Definition of the effective potential}

Based on the variables defined previously, we write the energy changes
around the RS, $\Delta E_{\rm eff}$ = $E_{\rm eff} - E_{\rm RS}$, as
\begin{equation}
\Delta E_{\rm eff} (\{\boldsymbol{u}_{i}\},\boldsymbol{\eta}) = E_{\rm
  p}(\{\boldsymbol{u}_{i}\}) + E_{\rm s}(\boldsymbol{\eta}) + E_{\rm
  sp}(\{\boldsymbol{u}_{i}\},\boldsymbol{\eta}) ,
\label{eq:energy_eff}
\end{equation}
where 
\begin{equation}
E_{\rm p}(\{\boldsymbol{u}_{i}\}) = E_{\rm
  har}(\{\boldsymbol{u}_{i}\}) + E_{\rm anh}(\{\boldsymbol{u}_{i}\})
\, .
\end{equation}
Here we use the subscript ``eff'' to distinguish between the energy
that our {\em effective} potential gives for configuration
$(\{\boldsymbol{u}_{i}\},\boldsymbol{\eta})$ and the {\em real} energy
$E(\{\boldsymbol{u}_{i}\},\boldsymbol{\eta})$ that we would obtain
from a first-principles simulation of the same atomic state. The above
terms are: the energy of the RS ($E_{\rm RS}$); the energy change when
the RS is distorted by atomic displacements $\{\boldsymbol{u}_{i}\}$
($E_{\rm p}$, where the ``p'' subscript stands for ``phonon''), which
we split into harmonic ($E_{\rm har}$) and anharmonic ($E_{\rm anh}$)
contributions; the energy change when we strain the RS ($E_{\rm s}$,
where the ``s'' subscript stands for ``strain''); and the additional
energy variations occurring when homogeneous strains and atomic
distortions appear simultaneously ($E_{\rm sp}$, where ``sp'' stands
for ``strain-phonon''). Let us discuss each of these terms.

\subsubsection{$E_{\rm har}(\{\boldsymbol{u}_{i}\})$ and
  $E_{\rm anh}(\{\boldsymbol{u}_{i}\})$}

Traditionally, the energy change caused by atomic distortions
$\{\boldsymbol{u}_{i}\}$ is written as a Taylor series around the RS
in the following way
\begin{equation}
\begin{split}
E_{\rm p}(\{\boldsymbol{u}_{i}\}) = & \, \frac{1}{2} \sum_{i\alpha
  j\beta} K^{(2)}_{i\alpha j\beta} u_{i\alpha}u_{j\beta} \, + \\
& \, \frac{1}{6} \sum_{i\alpha j\beta k\gamma} K^{(3)}_{i\alpha j\beta
  k\gamma} u_{i\alpha}u_{j\beta}u_{k\gamma} + ... \, ,
\end{split}
\label{eq:Eu-products}
\end{equation}
where the first line shows the harmonic terms included in $E_{\rm
  har}(\{\boldsymbol{u}_{i}\})$ and the second line gathers all the
higher order terms contained in $E_{\rm
  anh}(\{\boldsymbol{u}_{i}\})$. The tensor $\boldsymbol{K}^{(n)}$ is
formed by the $n$-th derivatives of the energy, with
\begin{equation}
K^{(n)}_{i\alpha j\beta ...} = \frac{\partial^{n} E_{\rm
    eff}}{\partial u_{i\alpha} \partial u_{j\beta} ...}\Bigr|_{\rm RS}
\, .
\end{equation}
Note that we assume that the RS is a stationary point of the PES
(i.e., a minimum or a saddle), so that $\boldsymbol{K}^{(1)} = 0$.

It is important to realize that the coefficients
$\boldsymbol{K}^{(n)}$ in Eq.~(\ref{eq:Eu-products}) are not
independent. At each order in the Taylor series, they are related by
the point and lattice-translational symmetries of the RS
structure. Additionally, and more fundamentally, they have to comply
with translational invariance in free space, which results in the
so-called {\sl acoustic sum rules} (ASRs). In essence, the ASRs
guarantee that a rigid translation of the material -- i.e., one given
by $u_{i\alpha}$ = $u_{\alpha}$, where $u_{\alpha}$ is an arbitrary
three-dimensional vector -- does not change the energy and does not
induce any forces on the atoms. To fulfill these conditions, the
$\boldsymbol{K}^{(n)}$ coefficients must satisfy
\begin{equation}
\sum_{i} K^{(n)}_{i\alpha j\beta k\gamma ...} = 0 \, , \;\;\; \forall
j,k,...,\alpha,\beta,\gamma,... \, ,
\label{eq:asr-gen}
\end{equation}
at all orders of the expansion. In the harmonic case with $n = 2$,
this reduces to the well-known ASR for the elements of the
force-constant matrix
\begin{equation}
\sum_{i} K^{(2)}_{i\alpha j\beta} = 0 \, , \;\;\; \forall
j,\alpha,\beta  \, .
\end{equation}
This set of conditions for the harmonic terms is rather manageable,
and allows for simple procedures to enforce the ASR in practice. For
example, a common strategy is to derive the self-energy parameters
from the interactions between different atoms, by taking
\begin{equation}
K^{(2)}_{i\alpha i\beta} = - \sum_{j\neq i} K^{(2)}_{i\alpha j\beta}
\, ,
\label{eq:ASR-transformation}
\end{equation}
and simultaneously imposing the symmetric character of the
force-constant matrix ($K^{(2)}_{i\alpha j\beta}$ = $K^{(2)}_{j\beta
  i\alpha}$). Note that such a {\em correction} is necessary whenever
we spatially truncate the interatomic couplings, as such an
approximation will generally break the ASR. Also, it is customary to
use this type of correction when dealing with a force-constant matrix
whose coefficients may suffer from some numerical noise or
inaccuracy. As we will discuss in Section~\ref{sec:par-calc}, we count
with well-established and widely-available first-principles methods to
compute a force-constant matrix that is ASR-compliant. Hence, we use
the above form [i.e., Eq.~(\ref{eq:Eu-products})] for the harmonic
term $E_{\rm har}$ in our models.

However, as one can imagine from Eq.~(\ref{eq:asr-gen}), enforcing the
ASR becomes much more intricate for $n>2$. In particular, it would
complicate enormously the procedure to compute the parameters in
$E_{\rm anh}$ discussed in Section~\ref{sec:par-calc}. Fortunately, in
that case we can resort to an alternative representation in which the
ASR is automatically satisfied at all orders.

Indeed, the energy $E_{\rm p}(\{\boldsymbol{u}_{i}\})$ can be
equivalently expanded as a function of {\sl displacement differences}
in the following way
\begin{widetext}
\begin{equation}
\begin{split}
E_{\rm p}(\{\boldsymbol{u}_{i}\}) = & \frac{1}{2}
\sum_{\substack{ijkh\\\alpha\beta}} \widetilde{K}^{(2)}_{ij\alpha
  kh\beta} (u_{i\alpha} - u_{j\alpha})(u_{k\beta} - u_{h\beta}) + \\ &
\frac{1}{6} \sum_{\substack{ijkhrt\\\alpha\beta\gamma}}
\widetilde{K}^{(3)}_{ij\alpha kh\beta rt\gamma} (u_{i\alpha} -
u_{j\alpha})(u_{k\beta} - u_{h\beta})(u_{r\gamma} - u_{t\gamma}) +
... \, .
\end{split}
\label{eq:Eu-differences}
\end{equation}
\end{widetext}
From this expression, it is obvious that $E_{\rm p}$ does not change
for a rigid displacement of the material, as every single term cancels
out in that case; it is also easy to prove that a rigid displacement
does not induce any forces on the atoms. Hence, the model parameters
$\widetilde{\boldsymbol{K}}^{(n)}$ do not need to satisfy any ASR to
guarantee translational invariance, which facilitates enormously the
task of fitting their values to best reproduce first-principles
results.\cite{fnaccuracy}

The relation between Eqs.~(\ref{eq:Eu-products}) and
(\ref{eq:Eu-differences}) is a subtle one and deserves a few
comments. ($i$) It is important to realize that these two expressions
for $E_{\rm p}$ are {\em not} connected by a simple transformation of
the basis in which we express the atomic distortions of the
material. Indeed, the atomic displacements $\{\boldsymbol{u}_{i}\}$ do
define the independent variables of our problem. In contrast, the set
of differences $\{(u_{i\alpha}-u_{j\alpha})\}$ has many more,
linearly-dependent members; hence, the displacement differences are
not an acceptable basis. ($ii$) It is possible to go from
Eq.~(\ref{eq:Eu-products}) to Eq.~(\ref{eq:Eu-differences}) by
application of the ASR at each order of the expansion. More precisely,
at a given order $n$, one can use the corresponding ASRs to write some
of the $K_{i\alpha j\beta ...}^{(n)}$ parameters as a function of the
rest, e.g. by performing substitutions as the one given in
Eq.~(\ref{eq:ASR-transformation}) for $n = 2$. The result of such a
procedure is an expression in terms of differences, as the one in
Eq.~(\ref{eq:Eu-differences}). However, there is no unique way to
perform such a transformation and, thus, the form of the resulting
energy function is somewhat arbitrary. Indeed, there are many ways in
which we can use the ASRs to rewrite Eq.~(\ref{eq:Eu-products}) [e.g.,
  for $n = 2$, Eq.~(\ref{eq:ASR-transformation}) is just one
  possibility among many others], and we see no clear reasons to
prefer any specific strategy. ($iii$) In
Eq.~(\ref{eq:Eu-differences}), it may look like we have {\em
  many-body} terms already at very low orders of the expansion. For
example, the harmonic terms can involve up to four different
atoms. Such couplings are the result of the ASR-related connections
between the energy derivatives $\boldsymbol{K}^{(n)}$, which result in
terms that look like many-body ones when we write the energy as a
function of displacement differences. ($iv$) It is possible to
understand better the inner structure of the difference terms of
Eq.~(\ref{eq:Eu-differences}). Thus, for example, for $n = 2$ it can
be seen that all four-body terms can be written as combinations of
two- and three-body terms, but three-body terms are in general not
reducible to two-body terms. These considerations are of little
importance for our present purposes, though, and we will not pursue
them further. ($v$) Finally, let us note that the
$\tilde{\boldsymbol{K}}^{(n)}$ parameters in
Eq.~(\ref{eq:Eu-differences}) can be viewed some sort of generalized
spring constants; this interpretation is especially apparent for the
pairwise terms involving products of the form
$(u_{i\alpha}-u_{j\alpha})^{n}$.

\subsubsection{$E_{\rm s}(\boldsymbol{\eta})$ and $E_{\rm
    sp}(\{\boldsymbol{u}_{i}\},\boldsymbol{\eta})$}

For the elastic energy $E_{\rm s}(\boldsymbol{\eta})$, we use a simple
Taylor series
\begin{equation}
E_{\rm s}(\boldsymbol{\eta}) = \frac{N}{2} \sum_{ab} C^{(2)}_{ab}
\eta_{a}\eta_{b} + \frac{N}{6} \sum_{abc} C^{(3)}_{abc}
\eta_{a}\eta_{b}\eta_{c} + ... \, ,
\label{eq:Es}
\end{equation}
where
\begin{equation}
C^{(m)}_{ab...} = \frac{1}{N} \frac{\partial^{m} E_{\rm eff}}{\partial
  \eta_{a} \partial\eta_{b} ...} \Bigr|_{\rm RS} \, ,
\end{equation}
and $N$ is the number of cells in the crystal. There is no linear term
in Eq.~(\ref{eq:Es}) because we assume that the RS is a stationary
point of the PES. The harmonic parameters in this series are the usual
elastic constants; more precisely, they are the so-called {\em
  frozen-ion} or {\em undressed} elastic constants, as they quantify
the elastic response of the material with the ions clamped at the
relative positions that they have in the RS.

For the strain-phonon interaction energy
$E_{sp}(\{\boldsymbol{u}_{i}\},\boldsymbol{\eta})$, we can write
\begin{widetext}
\begin{equation}
E_{\rm sp}(\{\boldsymbol{u}_{i}\},\boldsymbol{\eta}) = \frac{1}{2}
\sum_{a}\sum_{i\alpha} \Lambda^{(1,1)}_{ai\alpha} \eta_{a} u_{i\alpha}
+ \frac{1}{6} \sum_{a}\sum_{i\alpha j\beta} \Lambda^{(1,2)}_{ai\alpha
  j\beta} \eta_{a} u_{i\alpha} u_{j\beta} + \frac{1}{6}
\sum_{ab}\sum_{i\alpha} \Lambda^{(2,1)}_{abi\alpha} \eta_{a} \eta_{b}
u_{i\alpha} + ... \, .
\label{eq:Esp-products}
\end{equation}
\end{widetext}
The lowest-order coupling term $\boldsymbol{\Lambda}^{(1,1)}$
corresponds (except for non-essential prefactors) to the so-called
{\em force-response internal strain tensor}, and describes the forces
that act on the atoms as a consequence of homogeneous strains. Hence,
this kind of coupling contributes to determine the full, {\em
  relaxed-ion} or {\em dressed}, elastic response of the material, in
the way that is described e.g. in Ref.~\onlinecite{wu05}.

The $\boldsymbol{\Lambda}^{(m,n)}$ parameters in
Eq.~(\ref{eq:Esp-products}) have to comply with a set of ASRs that are
analogous to the ones discussed above for the $\boldsymbol{K}^{(n)}$
coefficients. As in the case of $E_{\rm p}$, we can use an alternative
expression for $E_{\rm sp}$, namely
\begin{widetext}
\begin{equation}
\begin{split}
E_{\rm sp}(\{\boldsymbol{u}_{i}\},\boldsymbol{\eta}) = \; & \frac{1}{2}
\sum_{a}\sum_{ij\alpha} \widetilde{\Lambda}^{(1,1)}_{aij\alpha}
\eta_{a} (u_{i\alpha} - u_{j\alpha}) + \frac{1}{6}
\sum_{a}\sum_{\substack{ijhk\\\alpha\beta}}
\widetilde{\Lambda}^{(1,2)}_{aij\alpha kh\beta} \eta_{a} (u_{i\alpha}
- u_{j\alpha}) (u_{k\beta}-u_{h\beta})\\
& + \frac{1}{6} \sum_{ab}\sum_{ij\alpha}
 \widetilde{\Lambda}^{(2,1)}_{abij\alpha} \eta_{a} \eta_{b}
 (u_{i\alpha} - u_{j\alpha}) + ... \, ,
\end{split}
\label{eq:Esp-differences}
\end{equation}
\end{widetext}
with $\widetilde{\boldsymbol{\Lambda}}^{(m,n)}$ parameters that are
free from ASR-related restrictions. Our choosing between the former or
the latter expressions for $E_{\rm sp}$ will be a matter of practical
convenience; more precisely, we will use the regular representation
[Eq.~(\ref{eq:Esp-products})] whenever we compute the parameters
directly from first principles, and the alternative one
[Eq.~(\ref{eq:Esp-differences})] in cases in which we need to fit the
parameters to reproduce specific first-principles results. This will
be discussed in detail in Section~\ref{sec:par-calc}.

\subsubsection{Symmetry considerations}

We will often deal with reference structures that present certain
lattice-translational and/or point symmetries. Such symmetries imply a
reduction in the number of independent parameters of the model, and we
can take advantage of them to simplify its construction. In the
following we describe the general ideas and procedures that one can
use to this end, resorting to the ABO$_{3}$ perovskites as a
convenient example in which the high symmetry of the RS (i.e., the
full cubic space group $Pm\bar{3}m$) results in great simplifications.

Let us denote a general symmetry operation by
$\{\boldsymbol{S}|\boldsymbol{t}\}$, where $\boldsymbol{S}$ is the
3$\times$3-matrix representation of a point symmetry and
$\boldsymbol{t}$ is a three-dimensional vector, both expressed in our
Cartesian reference. By applying such an operation to an arbitrary
vector $\boldsymbol{x}$, we obtain the transformed vector
$\boldsymbol{x}' = \{\boldsymbol{S}|\boldsymbol{t}\}\boldsymbol{x}$
given by
\begin{equation}
x'_{\alpha} = \sum_{\beta} S_{\alpha\beta} x_{\beta} + t_{\alpha} \, .
\end{equation}
For $\{\boldsymbol{S}|\boldsymbol{t}\}$ to be a symmetry of the RS, it
is necessary and sufficient to have
\begin{equation}
\{\boldsymbol{S}|\boldsymbol{t}\} (\boldsymbol{R}_{i} +
\boldsymbol{\tau}_{i}) = \boldsymbol{R}_{i'} + \boldsymbol{\tau}_{i'}
\, ,
\label{eq:symmetry_invariance}
\end{equation}
where, for any atom $i$, there is an atom $i'$ of the same atomic
species that satisfies this relation. In other words,
Eq.~(\ref{eq:symmetry_invariance}) states that the RS is invariant
upon the application of $\{\boldsymbol{S}|\boldsymbol{t}\}$.

The distortions of the RS transform as
\begin{equation}
u'_{i'\alpha} = \sum_{\beta} S_{\alpha\beta} u_{i\beta}
\, ,
\end{equation}
where $i$ and $i'$ are related by
Eq.~(\ref{eq:symmetry_invariance}), and we also have
\begin{equation}
\eta'_{\alpha\beta} = \sum_{\gamma\delta} S_{\alpha\gamma}
\eta_{\gamma\delta} (S^{-1})_{\delta\beta} = \sum_{\gamma\delta}
S_{\alpha\gamma} S_{\beta\delta} \eta_{\gamma\delta}\, ,
\end{equation}
where the strains are expressed in the Cartesian basis. Finally, the
symmetry condition for the energy reads
\begin{equation}
\Delta E_{\rm eff}(\{\boldsymbol{u}'_{i'}\},\boldsymbol{\eta}') =
\Delta E_{\rm eff}(\{\boldsymbol{u}_{i}\},\boldsymbol{\eta}) \, .
\end{equation}
Of course, similar relations hold for all the individual terms in the
energy and at all orders of the Taylor series [e.g., we have $E_{\rm
    anh}(\{\boldsymbol{u}'_{i'}\})$ = $E_{\rm
    anh}(\{\boldsymbol{u}_{i}\})$].

\begin{figure*}
\includegraphics[width=0.9\textwidth]{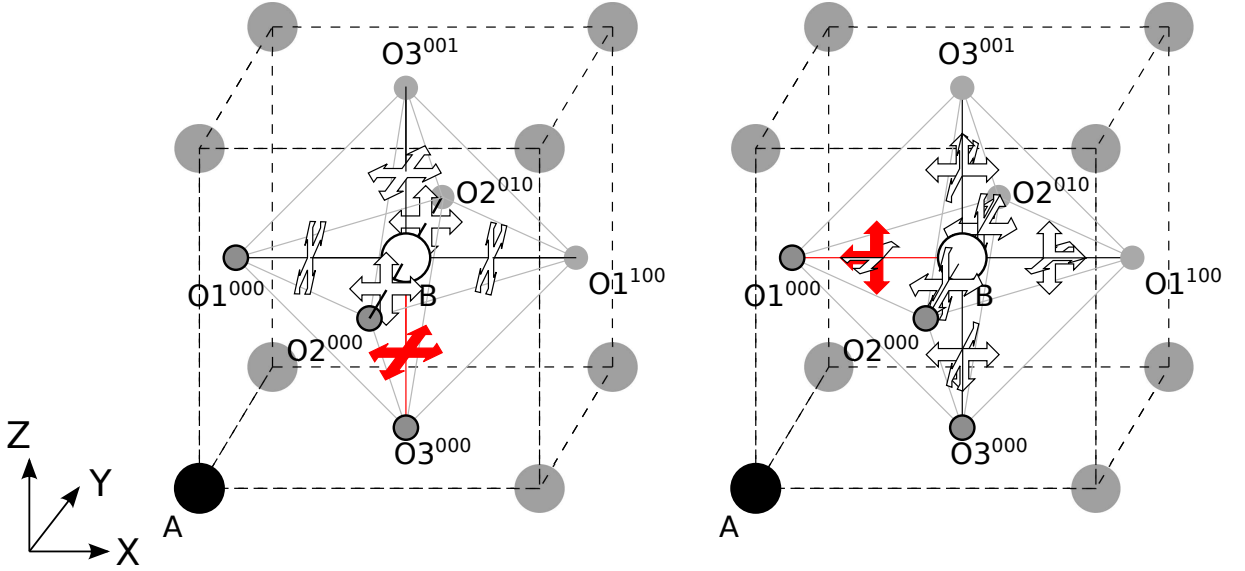}
\caption{(Color on line.) Example of symmetry adapted terms (SATs)
  involving displacements of neighboring B and O atoms. We show the
  SATs represented by the
  (B$_{x}-$O3$_{x}$)$^{2}$(B$_{y}-$O3$_{y}$)$^{2}$ (left) and
  (B$_{z}-$O1$_{z}$)$^{2}$(B$_{x}-$O1$_{x}$) (right) coupling terms
  (the representative terms are colored in the figure). For the atomic
  displacements we use the compact notation described in the text. An
  arrow along the $\alpha$ direction, and located at the center of the
  line connecting atoms $i$ and $j$, represents the
  $(u_{i\alpha}-u_{j\alpha})$ displacement difference. Whenever a
  displacement difference appears squared, we draw a double arrow,
  which indicates invariance under mirror-plane reflections. Without
  loss of generality, we assume that the B atom is located at the
  $l=0$ cell of the lattice [i.e., $\boldsymbol{R}_{l}$ =
    $\boldsymbol{R}_{0}$ = $(0,0,0)$]; superscripts at the oxygen
  sites denote the cell to which they correspond.}
\label{fig:SATs}
\end{figure*}

Let us describe how these general symmetry relations allow us to
simplify our model potential. Given a particular product of
$u_{i\alpha}$ displacements and $\eta_{a}$ strains in the Taylor
series, we can use the operations of the space group to generate the
collection of symmetry-related products, which will involve
transformed $u'_{i'\alpha}$ and $\eta'_{a}$
distortions. Figure~\ref{fig:SATs} illustrates this process
pictorially. For example, in the left panel we start with the product
$(u_{0{\rm B}x}-u_{0{\rm O3}x})^{2} (u_{0{\rm B}y}-u_{0{\rm
    O3}y})^{2}$ that couples atoms B and O3 in the unit cell at the
origin of our coordinate system [i.e., without loss of generality, we
  choose $\boldsymbol{R}_{l}$ = $\boldsymbol{R}_{0}$ =
  $(0,0,0)$]. Then, by application of the symmetry operations of the
cubic space group of the ideal perovskite structure, we can generate a
collection of related products; for example, a 90$^{\circ}$ rotation
about the $y$ axis transforms the original product into $(u_{0{\rm
    B}z}-u_{0O1z})^{2} (u_{0{\rm B}y}-u_{0O1y})^{2}$,
etc. Figure~\ref{fig:SATs} sketches the products thus generated and
involving the B atom in the $l=0$ cell; lattice-translational symmetry
leads to analogous couplings {\em centered} at all other B atoms in
the crystal.

Naturally, these symmetry-related couplings must contribute to the
energy in a very specific way. Continuing with the above example, the
couplings {\em represented} by $(u_{0{\rm B}x}-u_{0{\rm O3}x})^{2}
(u_{0{\rm B}y}-u_{0{\rm O3}y})^{2}$ appear in our potential in the
form
\begin{widetext}
\begin{equation}
\begin{split}
&\; \widetilde{K}^{(4)}_{0{\rm B}0{\rm O3}x,0{\rm B}0{\rm O3}x,0{\rm
    B}0{\rm O3}y,0{\rm B}0{\rm O3}y} \Bigl[ (u_{0{\rm B}x}-u_{0{\rm
      O3}x})^2(u_{0{\rm B}y}-u_{0{\rm O3}y})^2 + (u_{0{\rm
      B}z}-u_{0{\rm O1}z})^2(u_{0{\rm B}y}-u_{0{\rm O1}y})^2 + ... \;
  \Bigr] \\
& \equiv \; \widetilde{K}_{\#15} \Bigl[
({\rm B}_{x}-{\rm O}3_{x}^{000})^{2}({\rm B}_{y}-{\rm O}3_{y}^{000})^{2} +
({\rm B}_{x}-{\rm O}3_{x}^{001})^{2}({\rm B}_{y}-{\rm O}3_{y}^{001})^{2} +
({\rm B}_{x}-{\rm O}2_{x}^{000})^{2}({\rm B}_{z}-{\rm
    O}2_{z}^{000})^{2} \\
& \;\;\;\;\;\; + ({\rm B}_{x}-{\rm O}2_{x}^{010})^{2}({\rm B}_{z}-{\rm
    O}2_{z}^{010})^{2} + ({\rm B}_{z}-{\rm O}1_{z}^{000})^{2}({\rm
    B}_{y}-{\rm O}1_{y}^{000})^{2} + ({\rm B}_{z}-{\rm
    O}1_{z}^{100})^{2}({\rm B}_{y}-{\rm O}1_{y}^{100})^{2} \Bigr] \, ,
\end{split}
\end{equation}
\end{widetext}
where $\widetilde{K}_{\#15}$ is the name we use for this specific
coupling parameter in Table~\ref{tab:SATs}. Here we have introduced a
compact notation, so that we denote by B$_{\alpha}$ the displacements
$u_{0{\rm B}\alpha}$ of the central B atom, by
O1$_{\alpha}^{n_{l1}n_{l2}n_{l3}}$ the displacements $u_{l{\rm
    O1}\alpha}$ of the O1 atom at cell $l$, etc. ($\boldsymbol{R}_{l}$
is defined by the integers $n_{l1}$, $n_{l2}$, and $n_{l3}$ as in the
caption of Fig.~\ref{fig:model-perovskite}.) This is what we call a
{\em symmetry-adapted term} (SAT), which is fully specified by one
representative coupling and its associated parameter.

\begin{table}
\caption{Representatives of the symmetry-adapted terms (SATs) that
  couple first-nearest neighbors in the ABO$_3$ cubic structure. The
  atom labels correspond to those in
  Fig.~\protect\ref{fig:SATs}. (Note that all these representative
  couplings can be chosen so that the two atoms involved are in the
  same crystal cell.) For the atomic displacements we use the compact
  notation described in the text. We number the couplings to refer to
  them easily in the text. This also allows for a compact notation for
  the coupling coefficients; for example, the SATs sketched in
  Fig.~\protect\ref{fig:SATs} correspond to coefficients
  $\widetilde{K}_{\#15}$ (left) and $\widetilde{K}_{\#11}$ (right). }

\vskip 2mm

\begin{tabular*}{0.7\columnwidth}{@{\extracolsep{\fill}}cl}
\hline\hline\\
\multicolumn{2}{l}{Third-order A-O terms} \\
1&
(A$_{z}$$-$O1$_{z}$)$^{3}$\\
2&
(A$_{z}$$-$O3$_{z}$)$^{2}$ (A$_{x}$$-$O3$_{x}$) \\
3&
(A$_{x}$$-$O2$_{x}$)$^{2}$ (A$_{z}$$-$O2$_{z}$)\\ 
\\
\multicolumn{2}{l}{Fourth-order A-O terms} \\
4&
(A$_{y}$$-$O2$_{y}$)$^{4}$\\
5&
(A$_{z}$$-$O2$_{z}$)$^{4}$\\
6&
(A$_{z}$$-$O3$_{z}$)$^{2}$ (A$_{y}$$-$O3$_{y}$)$^{2}$\\
7&
(A$_{z}$$-$O2$_{z}$)$^{2}$ (A$_{x}$$-$O2$_{x}$)$^{2}$\\ 
8&
(A$_{y}$$-$O1$_{y}$)$^{3}$ (A$_{z}$$-$O1$_{z}$)\\
9&
(A$_{x}$$-$O1$_{x}$)$^{2}$ (A$_{y}$$-$O1$_{y}$) (A$_{z}$$-$O1$_{z}$)\\
\\
\multicolumn{2}{l}{Third-order B-O terms} \\
10&
(B$_{x}$$-$O1$_{x}$)$^{3}$\\
11&
(B$_{z}$$-$O1$_{z}$)$^{2}$ (B$_{x}$$-$O1$_{x}$) \\
\\
\multicolumn{2}{l}{Fourth-order B-O terms} \\
12&
(B$_{x}$$-$O1$_{x}$)$^{4}$\\
13&
(B$_{z}$$-$O1$_{z}$)$^{4}$\\
14&
(B$_{x}$$-$O3$_{x}$)$^{2}$ (B$_{z}$$-$O3$_{z}$)$^{2}$\\
15&
(B$_{x}$$-$O3$_{x}$)$^{2}$ (B$_{y}$$-$O3$_{y}$)$^{2}$ \\
\\
\hline\hline
\end{tabular*}
\label{tab:SATs}
\end{table}

\begin{table}
\caption{Same as in Table~\protect\ref{tab:SATs}, but involving
  couplings between strains (linear) and atomic displacements of
  nearest-neighboring atoms (quadratic). Strains are given in the
  Cartesian notation $\eta_{\alpha\beta}$ to facilitate the
  interpretation of the terms.}

\vskip 2mm

\begin{tabular*}{0.6\columnwidth}{@{\extracolsep{\fill}}cl}
\hline\hline\\
\multicolumn{2}{l}{A-O terms} \\
1&$({\rm A}_{y}-{\rm O}2_{y})^{2}$ $\eta_{yy}$\\
2&$({\rm A}_{z}-{\rm O}2_{z})^{2}$ $\eta_{zz}$\\
3&$({\rm A}_{z}-{\rm O}3_{z})^{2}$ $\eta_{yy}$\\
4&$({\rm A}_{x}-{\rm O}2_{x})^{2}$ $\eta_{xx}$\\
5&$({\rm A}_{z}-{\rm O}1_{z})^{2}$ $\eta_{yy}$\\
6&$({\rm A}_{x}-{\rm O}1_{x})^{2}$ $\eta_{zy}$\\
7&$({\rm A}_{y}-{\rm O}1_{y})^{2}$ $\eta_{zy}$\\
8&$({\rm A}_{x}-{\rm O}2_{x})$ $({\rm A}_{z}-{\rm O}2_{z})$ $\eta_{yy}$\\
9&$({\rm A}_{y}-{\rm O}2_{y})$ $({\rm A}_{x}-{\rm O}2_{x})$ $\eta_{yx}$\\
10&$({\rm A}_{z}-{\rm O}3_{z})$ $({\rm A}_{y}-{\rm O}3_{y})$ $\eta_{xz}$\\
11&$({\rm A}_{x}-{\rm O}3_{x})$ $({\rm A}_{y}-{\rm O}3_{y})$ $\eta_{yy}$\\
12&$({\rm A}_{y}-{\rm O}3_{y})$ $({\rm A}_{x}-{\rm O}3_{x})$ $\eta_{yx}$\\ \\

\multicolumn{2}{l}{B-O terms} \\
13&$({\rm B}_{z}-{\rm O}3_{z})^{2}$ $\eta_{zz}$\\
14&$({\rm B}_{y}-{\rm O}3_{y})^{2}$ $\eta_{zz}$\\
15&$({\rm B}_{y}-{\rm O}2_{y})^{2}$ $\eta_{xx}$\\
16&$({\rm B}_{z}-{\rm O}2_{z})^{2}$ $\eta_{zz}$\\
17&$({\rm B}_{x}-{\rm O}2_{x})^{2}$ $\eta_{zz}$\\
18&$({\rm B}_{x}-{\rm O}3_{x})$ $({\rm B}_{y}-{\rm O}3_{y})$ $\eta_{xy}$\\ 
19&$({\rm B}_{z}-{\rm O}1_{z})$ $({\rm B}_{x}-{\rm O}1_{x})$ $\eta_{zx}$\\ \\

\hline\hline
\end{tabular*}
\label{tab:etaSATs}
\end{table}

Note also that, in some cases, by applying all the space group
operations to a representative coupling we may generate a SAT that
exactly cancels out. Hence, working with SATs provides us with an
automatic way to identify couplings that are forbidden by symmetry,
which may result in drastic simplifications of our model
potentials. For example, in the case of our ABO$_{3}$ perovskites, the
symmetry of the RS guarantees that all the bilinear strain-phonon
couplings are zero (i.e., $\boldsymbol{\Lambda}^{(1,1)}$ =
$\widetilde{\boldsymbol{\Lambda}}^{(1,1)}$ = 0), a fact that has
actual physical consequences on the response properties of the cubic
phase of such materials.

In our investigation of ABO$_{3}$ perovskites, we always worked with
SATs. This is clearly the recommended strategy to follow: implementing
the automatic generation of the SATs is relatively easy (by systematic
application of the RS symmetries as outlined above) and results in
more transparent and easier-to-construct models. Thus, we will refer
to SATs when describing the effective models for ABO$_{3}$ compounds;
the relevant ones (i.e., their representative couplings) are listed in
Tables~\ref{tab:SATs} and \ref{tab:etaSATs}.

\subsubsection{Long-range interactions in insulators \label{sec:long-range}}

The potentials described above can in principle involve interatomic
interactions of arbitrary spatial range. However, in practice we will
truncate the spatial extent of such interactions, which will
constitute one of the approximations in our models. Generally
speaking, such a truncation can be expected to work well in metals,
where the free charges provide an efficient means of screening. In
contrast, the truncation is not justified when we deal with
semiconductors or insulators, where long-range (strictly speaking,
infinite-range) Coulomb interactions must necessarily be
considered.\cite{fnsilicon} Fortunately, such couplings have a
well-known analytic form in the limit of long distances, and they can
be conveniently treated in a way that is essentially exact.

To understand the role of ion-ion Coulomb interactions in insulators,
let us consider two separate effects. (In the following we will
implicitly consider the case of short-circuit boundary conditions,
which corresponds to the ideal situation for an infinite bulk
material. The treatment of different electrostatic boundary conditions
is discussed e.g. in Ref.~\onlinecite{stengel09}.) First, these
interactions give raise to the so-called {\em Madelung field} that
contributes to determine the cohesive energy of the material. In our
model potentials, such a Madelung field is captured in the energy of
the RS. Hence, by taking $E_{\rm RS}$ directly as a result of the
first-principles calculations, we avoid the need to model the Madelung
energy, as well as the other effects (e.g., atomic and short-range
interactions associated with chemical bonding) that control the basic
cohesive energy. Second, the Coulombic interaction between ions also
influences the energy changes associated with the distortions of the
RS. To leading order in the Taylor series, such an effect is
essentially captured by the electrostatic interaction between the {\em
  dipoles} that appear when ions move from their RS positions. Such
atomic dipoles are usually written, within a linear approximation, as
\begin{equation}
d_{i\alpha} = \sum_{\beta} Z^{*}_{i\beta\alpha} u_{i\beta} \, ,
\label{eq:dipole}
\end{equation}
where $\boldsymbol{Z}^{*}_{i}$ is the so-called {\em Born
  effective-charge tensor} or {\em dynamical-charge tensor} for atom
$i$. (Strictly speaking, we should talk about dipole {\em
  differences}. Yet, here we will assume that these local dipoles are
zero in the RS, which will be the natural choice in most cases.) Note
that the Born charge $Z^{*}_{i\beta\alpha}$ quantifies the dipole
caused by the displacement of the ionic charge associated with ion $i$
at its RS position, as well as all the additional effect arising from
the electronic rearrangement that occurs in response to the atomic
distortion. In the case of insulating ABO$_{3}$ perovskites like
PbTiO$_3$ and SrTiO$_3$, the electronic effects are very large and
result in Born charges that even double the value corresponding to the
rigid-ion limit.\cite{zhong94b} Such huge dynamical charges reflect
changes in the oxygen--cation bonding that play a crucial role in the
ferroelectric and response properties of those
materials.\cite{posternak94}

Hence, when working with insulators, it will be convenient to split
the energy terms involving atomic distortions $\boldsymbol{u}_{i}$
into short-range (``sr'') and long-range (``lr'') parts. Thus, for
example, we have
\begin{equation}
\boldsymbol{K}^{(n)} = \boldsymbol{K}^{(n),sr} +
\boldsymbol{K}^{(n),lr}
\end{equation}
for the couplings in $E_{\rm p}$, where it is important to note that
the decomposition can be done at all orders in the Taylor
series. Analogously, the strain-phonon terms in $E_{\rm sp}$ can be
split as
\begin{equation}
\boldsymbol{\Lambda}^{(m,n)} = \boldsymbol{\Lambda}^{(m,n),sr} +
\boldsymbol{\Lambda}^{(m,n),lr} \, .
\end{equation}
Of course, analogous splittings can be considered for the parameters
that appear in our displacement-difference representation of
Eqs.~(\ref{eq:Eu-differences}) and (\ref{eq:Esp-differences}).

Here we will only discuss the lowest-order dipole-dipole interactions,
which are captured by the harmonic couplings $\boldsymbol{K}^{(2)}$
and have been described in detail in the literature. (Harmonic
couplings involving other terms in the multipole expansion of the
electrostatic energy also exist; as usually done in first-principles
treatments, we will neglect their contribution to the long-range part
of the energy, and effectively capture their possible effects in the
short-range part.) Following Gonze and Lee,\cite{gonze97} we write the
long-range couplings as
\begin{equation}
\begin{split}
&K^{(2),lr}_{i\alpha j\beta} = \\ 
&\sum_{\gamma\delta} Z^{*}_{i\alpha\gamma} Z^{*}_{j\alpha\delta}
  \left(\frac{(\boldsymbol{\epsilon}_{\infty}^{-1})_{\gamma\delta}}{D^3}
  - \frac{3\Delta_{\gamma}\Delta_{\delta}}{D^5}\right)
  (\det{\boldsymbol{\epsilon}_{\infty}})^{-1/2} \, ,
\end{split}
\label{eq:lr}
\end{equation}
where
\begin{equation}
\Delta_{\alpha} = \sum_{\beta}
(\boldsymbol{\epsilon}_{\infty}^{-1})_{\alpha\beta} \Delta r_{\beta}
\end{equation}
and
\begin{equation}
D  =  \sqrt{\boldsymbol{\Delta}\cdot\Delta\boldsymbol{r}} \, ,
\end{equation}
with
\begin{equation}
\Delta\boldsymbol{r} = \boldsymbol{R}_{j} + \boldsymbol{\tau}_{j} -
\boldsymbol{R}_{i} - \boldsymbol{\tau}_{i} \, .
\end{equation}
This is the usual expression for the Coulombic interaction between two
dipoles, generalized for a medium that presents an arbitrary
dielectric tensor $\boldsymbol{\epsilon}_{\infty}$ quantifying the
purely electronic (frozen-ion) response of the material. As discussed
by Gonze and Lee,\cite{gonze97} Eq.~(\ref{eq:lr}) captures the
non-analytical behavior of the phonon bands for homogeneous
($\boldsymbol{q} = 0$) distortions, and the related electrostatic
effects (e.g., the so-called longitudinal-optical--transversal-optical
splitting of the phonon frequencies). It is also trivial to show that
the ASR for the $\boldsymbol{K}^{(2),lr}$ coefficients translates into
the condition
\begin{equation}
\sum_{i} Z^{*}_{i\alpha\beta} = N \sum_{\kappa}
Z^{*}_{\kappa\alpha\beta} = 0 \;\;\; \forall
\alpha,\beta \;\; ,
\label{eq:ZASR}
\end{equation}
which guarantees that no net dipole is created by a rigid displacement
of all the atoms in the crystal. Note that the
$\boldsymbol{Z}^{*}_{i}$ tensors are cell-independent, which allows us
to use the notation $\boldsymbol{Z}^{*}_{\kappa}$. Finally, let us
mention that in an actual atomistic simulation, which usually involves
a periodically-repeated simulation box or supercell, such
infinitely-ranged couplings can be accurately computed by performing
an Ewald summation, as described e.g. in Ref.~\onlinecite{zhong95a}.

In this work we only considered the Coulombic dipole-dipole term
associated with $\boldsymbol{K}^{(2)}$. Indeed, as discussed below,
higher-order long-ranged couplings in $E_{\rm p}$, and further
interactions involving strain in $E_{\rm sp}$, were either neglected
or treated in an effective way. These approximations, which follow the
spirit of the usual effective-Hamiltonian approach to perovskite
oxides, will be discussed in Sections~\ref{sec:long-range-pars} and
\ref{sec:application}.

\subsubsection{Miscellaneous remarks}

We conclude this Section by commenting on various aspects of the model
potentials just described.

{\sl Approximations involved}.-- Typically, to construct an effective
potential for a material, one starts by considering the simplest
possible model that makes physical sense, and then extends it only as
much as needed to get a sufficiently accurate description of the
first-principles data of interest (i.e., a good description of what is
usually called the {\em training set} of first-principles
results). Given the conceptual simplicity of our proposed potentials,
it is straightforward to identify three qualitatively different ways
in which they can be systematically extended. Indeed, our models can
be improved as regards ($i$) the order of the polynomial expansion,
($ii$) the spatial range of the interatomic couplings considered, and
($iii$) the {\em complexity} of the coupling terms, i.e., the maximum
number of atoms (bodies) involved in the couplings. These three
truncations constitute the approximations of our models.

{\sl Relation with effective-Hamiltonian work}.-- For the most part,
the connections between our method and the above-mentioned
effective-Hamiltonian approach are rather obvious. Yet, there are a
couple of subtle points that deserve a comment.

The effective Hamiltonians often include local variables that account
for the {\em inhomogeneous} strains that may occur in the material;
further, the energy landscape for such local strains is typically
derived from the elastic constants associated with the homogeneous
ones, following the approximation proposed by Keating.\cite{keating66}
In our models, inhomogeneous strains are naturally captured by the
appropriate atomic distortions $\{\boldsymbol{u}_{i}\}$, as
illustrated in Figs.~\ref{fig:strain-decomp}(c) and
\ref{fig:strain-decomp}(d). The energy changes associated with such
local strains are given by $E_{\rm p}$, and there is no need to derive
them from the elastic constants for homogeneous cell deformations. (Of
course, one should note that the force constants
$\boldsymbol{K}^{(2)}$ and the elastic constants
$\boldsymbol{C}^{(2)}$ are connected by well-known relations, and the
latter can be computed from knowledge of the
former.\cite{born-book1954} As explained in
Section~\ref{sec:par-calc}, we include in our models the exact
first-principles results for both $\boldsymbol{K}^{(2)}$ and
$\boldsymbol{C}^{(2)}$, so that the relations between such
coefficients are fulfilled by construction.) Additionally, our models
also capture correctly the energy changes associated with strong
strain gradients. This is a definite improvement over the usual
effective-Hamiltonian approach, especially when taking into account
the growing interest in flexoelectric effects resulting from large
strain gradients near ferroelectric domain walls,\cite{catalan11} etc.

Secondly, the action of an external electric field $\boldsymbol{\cal
  E}$ can be trivially incorporated in an effective-Hamiltonian
simulation by including the leading coupling term between the field
and the local dipoles that are the basic variables of the
model.\cite{garcia98} Equivalently, within our approach (and as long
as we are dealing with insulators), we can use the effective-charge
tensors $\boldsymbol{Z}^{*}_{i}$ to compute the local dipole
$\boldsymbol{d}_{i}$ resulting from atomic displacements
[Eq.~(\ref{eq:dipole})], and write the corresponding energy as
\begin{equation}
E_{\rm eff}(\{\boldsymbol{u}_i\},\boldsymbol{\eta};\boldsymbol{\cal
  E}) = E_{\rm eff}(\{\boldsymbol{u}_i\},\boldsymbol{\eta}) -
\sum_{i\alpha} d_{i\alpha} {\cal E}_{\alpha} \, .
\end{equation}
Finally, let us note that the action of an external stress or pressure
$\boldsymbol{\sigma}$ can be treated in an analogous
way,\cite{garcia98} by introducing
\begin{equation}
E_{\rm
  eff}(\{\boldsymbol{u}_i\},\boldsymbol{\eta};\boldsymbol{\sigma}) =
E_{\rm eff}(\{\boldsymbol{u}_i\},\boldsymbol{\eta}) + N \sum_{a}
\eta_{a} \sigma_{a} \, .
\end{equation}
Here the sign convention is chosen so that a positive stress implies a
compression of the material.

{\sl Implementation in a simulation code \label{sec:practical}}.-- Let
us briefly mention some details of our implementation of a statistical
simulation [in particular, a Monte Carlo (MC) scheme] based on our
model potentials.

First, let us note that the long-range part of the force constant
matrix, $\boldsymbol{K}^{(2),lr}$, depends on the specific size and
shape of the periodically-repeated simulation box used for the MC
runs. Hence, for a given supercell, we compute these parameters before
the MC simulation starts, by performing the corresponding Ewald sums
that take into account interactions between periodically-repeated
dipole images. Then, we add up the long-range and short-range parts of
$\boldsymbol{K}^{(2)}$ to obtain a total harmonic interaction term
that effectively couples all atoms in the simulation box. This is what
we use for the energy evaluations in the simulation.

Once we have a supercell-dependent potential, the underlying
lattice-translational symmetry allows us to store only the
interactions between the atoms in one elemental unit cell and all
other atoms in the supercell. Hence, the storage requirements grow
linearly with number of unit cells in the supercell.

In our MC simulations, we attempt to change the strains only after
completing one sweep through all the atoms in the simulation
supercell. It is therefore convenient to recalculate the parameters
controlling the energetics of the atomic displacements, such as for
example
\begin{equation}
K^{(2)}_{i\alpha j\beta}\Bigr|_{\boldsymbol{\eta}} = K^{(2)}_{i\alpha
  j\beta} + \sum_{a} \Lambda^{(1,2)}_{a i\alpha j\beta} \eta_{a} +
\mathcal{O}(\eta^{2}) \, ,
\end{equation}
after the strains are updated. These strain-dependent parameters are
then used for energy evaluations during the sweep over atomic
displacements.

The SATs for the calculation of the anharmonic part of our models are
automatically generated based on the symmetry of the RS. We store them
in symbolic form, so that they can be used both for the calculation of
energy and ({\sl via} a simple manipulation of the polynomial) forces
on the atoms.

\subsection{Parameter Calculation \label{sec:par-calc}}

Once we have defined a potential, many schemes can be applied to
calculate its parameters. Here we describe the strategy that we
followed in this first application of our effective models, which
takes advantage of the direct availability of first-principles results
for many of the terms in the potential. Some approximations that we
used for the treatment of long-range interactions, which are somewhat
specific to the case of insulators undergoing structural phase
transitions, are also described.

\subsubsection{Parameters computed directly from first principles}

The low-order couplings of our model potentials quantify the response
of the RS of the material to small perturbations, may they be atomic
distortions, cell strains, or a combination of both. In particular,
the leading harmonic terms $\boldsymbol{K}^{(2)}$,
$\boldsymbol{C}^{(2)}$, and $\boldsymbol{\Lambda}^{(1,1)}$ can be
obtained directly from density-functional perturbation theory (DFPT)
calculations as those described in Refs.~\onlinecite{gonze97},
\onlinecite{baroni01} and \onlinecite{wu05}. DFPT schemes are
efficiently implemented in widely available first-principles codes,
such as the {\sc Abinit} package\cite{abinit09} used in this
work. Alternatively, one could obtain the same information by
performing systematic finite-difference calculations considering both
atomic displacements and strains. Such an approach, which is somewhat
more elementary but equally valid, is available in all major
first-principles packages. Hence, we can conclude that computing {\em
  exactly} the harmonic parameters of models like ours is a trivial
task nowadays.

\begin{figure}[t!]
\includegraphics[width=\columnwidth]{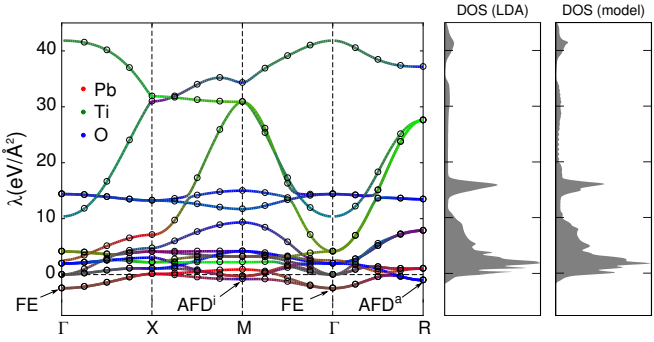}
\caption{(Color on line.) Left: Dispersion bands of cubic PbTiO$_3$,
  as calculated from first principles (lines) and obtained from our
  effective model (circles). The bands correspond to the eigenvalues
  $\lambda_{\boldsymbol{q}j}$ of the Fourier-transformed
  force-constant matrices $\boldsymbol{K}^{(2)}_{\boldsymbol{q}}$,
  which we call stiffness coefficients. The leading structural
  instabilities are labeled (see the text), and sketched in
  Fig.~\protect\ref{fig:instabilities}. The color code indicates the
  dominant atomic character of the
  $\boldsymbol{K}^{(2)}_{\boldsymbol{q}}$ eigenvectors. Right: Density
  of states (DOS) plots constructed from the
  $\boldsymbol{K}^{(2)}_{\boldsymbol{q}}$ eigenvalues, as obtained
  from first-principles simulations using a very fine $\boldsymbol{q}$
  point mesh, and from our effective potential by solving the
  eigenmode problem for an 8$\times$8$\times$8 supercell and making
  use of a simple interpolation between the computed eigenvalues.}
\label{fig:phonons-pto}
\end{figure}

\begin{figure}[t!]
\includegraphics[width=0.9\columnwidth]{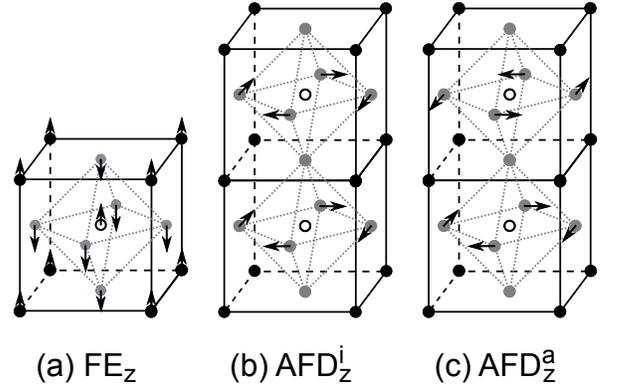}
\caption{Sketch of the atomic displacements corresponding to the most
  important structural instabilities in ABO$_{3}$ perovskite
  oxides. Panel~(a): Ferroelectric instability. Panel~(b):
  Anti-ferrodistortive instability with neighboring O$_{6}$ octahedra
  along the $z$ direction rotating in phase. Panel~(c): Same as in (b),
  but with octahedral rotations modulated in anti-phase along $z$.}
\label{fig:instabilities}
\end{figure}

Let us stress that the ability to incorporate an exact description of
the harmonic energy of the material {\em by construction} is a great
asset of our models. Indeed, in most materials the thermodynamic
properties are essentially captured at the harmonic level, with small
corrections coming from anharmonic effects; hence, a good description
of the harmonic lattice-dynamical properties is critical. Further,
even in cases with soft-mode-driven phase transitions, it is the
harmonic part of the energy what essentially determines the nature of
the leading structural instabilities. Hence, also in such situations,
a faithful harmonic description seems mandatory to have an accurate
model. Figure~\ref{fig:phonons-pto} shows representative results for
our model of PbTiO$_{3}$. As we can see, the description of the
force-constant bands of the RS is exact, and the small discrepancies
between the shown density-of-states (DOS) plots come from differences
in the way BZ integrations are performed in {\sc Abinit} and in our
codes. The most important structural instabilities, marked in
Fig.~\ref{fig:phonons-pto} and sketched in
Fig.~\ref{fig:instabilities}, are also reproduced exactly.

As regards the anharmonic terms, one could try a similar direct
calculation of each one of the parameters. For example, to compute the
strain-phonon couplings $\boldsymbol{\Lambda}^{(1,2)}$, one could run
DFPT calculations for the RS subject to a small strain
$\delta\boldsymbol{\eta}$. The resulting force-constant matrix would
be described in our model by
\begin{equation}
K^{(2)}_{i\alpha j\beta}\Bigr|_{\delta\boldsymbol{\eta}} =
K^{(2)}_{i\alpha j\beta} + \sum_{a} \Lambda^{(1,2)}_{a i\alpha j\beta}
\delta\eta_{a} \, ,
\label{eq:finitediff-lambda}
\end{equation}
which would allow us to calculate the targeted couplings. Following a
similar scheme -- e.g., by running DFPT calculations of distorted
configurations in which some atomic displacements are frozen in -- one
could access the parameters in $E_{\rm anh}$.

As described below, we tried such an approach when constructing our
models for PbTiO$_{3}$ and SrTiO$_{3}$, specifically in what regards
the strain-phonon couplings. Based on our experience, we believe that
such a systematic scheme may render accurate potentials in relatively
simple cases, i.e., whenever the RS does not present structural
instabilities. On the other hand, in the challenging situations here
considered, this strategy may be impractical if a very precise
description of some PES features is targeted. Indeed, we found that
the PES of materials like PbTiO$_{3}$ or SrTiO$_{3}$ is strongly
anharmonic; more precisely, if we aimed at an accurate description of
the {\em whole} PES connecting the RS with the lower-energy phases, we
would need to consider a Taylor series extending up to a rather high
order. In such cases it seems more convenient to adopt an effective
approach, aiming at reproducing the PES {\em only around} the RS and
the most relevant low-energy structures. This permits a lower-order
expansion that quantitatively captures the main effects and retains
much of the physical transparency of the simpler
(effective-Hamiltonian and phenomenological) models traditionally used
to investigate phase transitions, which include only as many terms as
strictly needed for a qualitatively correct description.

\subsubsection{Parameters fitted to first-principles
  results \label{sec:fit-pars}} 

To compute the higher-order couplings of our effective potentials --
i.e., $\tilde{\boldsymbol{K}}^{(n)}$ with $n > 2$ and
$\tilde{\boldsymbol{\Lambda}}^{(m,n)}$ with $m+n > 2$ --, it is
convenient to implement a fitting procedure aimed at obtaining a model
that reproduces a {\em training set} of first-principles results. Here
we describe the strategy we adopted in our work with PbTiO$_3$ and
SrTiO$_3$, where the training set was composed of low-energy
structures that are more stable than the RS, and the key properties
that we request our models to capture are energy differences and
equilibrium atomic configurations. Nevertheless, the ideas presented
are rather general and can be easily adapted to other situations.

In essence, our parameter-optimization calculations were based on
three {\em goal functions} defined in the following way. Let the
superindex $s$ number the structures
$(\{\boldsymbol{u}_{i}^{s}\},\boldsymbol{\eta}^{s})$ in our training
set. First, to get our model to reproduce the first-principles
energies $\{E^{s}\}$, we considered the goal function
\begin{equation}
 \mathcal{GF}_{E}(\mathcal{P}) = \sum_{s} \left[ E_{\rm
     eff}[\mathcal{P}](\{\boldsymbol{u}_{i}^{s}\},\boldsymbol{\eta}^{s})
   - E^{s}\right]^{2} \, ,
\label{eq:GFE}
\end{equation}
where ${\mathcal P}$ represents all the {\em free} adjustable
coefficients in the model and the parametric dependence of $E_{\rm
  eff}$ on ${\mathcal P}$ is indicated. Second, all the structures in
our training sets were stationary points of the PES (minima or
saddles). Hence, we imposed the zero-gradient condition for such
structures by minimizing the goal function
\begin{equation}
 \mathcal{GF}_{\nabla E}(\mathcal{P}) = \sum_{s} \left\| \nabla E_{\rm
   eff}[\mathcal{P}](\{\boldsymbol{u}_{i}^{s}\},\boldsymbol{\eta}^{s})
 \right\|^{2} \, ,
\label{eq:GFDE}
\end{equation}
where the gradient includes derivatives with respect to both atomic
distortions and cell strains. Finally, aiming at an improved
description of the lattice-dynamical properties of key low-energy
structures, we also used a goal function that contains information
about the corresponding Hessian matrices. More precisely, we used
\begin{equation}
 \mathcal{GF}_{\rm hess}({\mathcal P}) = \sum_{s} \sum_{\boldsymbol{q}
   \in \{\boldsymbol{q}\}_{s}} \mathcal{D}^{s}[{\mathcal
     P}](\boldsymbol{q}) \, ,
\label{eq:GFphon}
\end{equation}
where $\{\boldsymbol{q}\}_{s}$ is a set of $q$-points of the first
Brillouin zone of structure $s$ (we restricted ourselves to
zone-center and zone-boundary $q$-points). The function ${\mathcal
  D}^{s}[{\mathcal P}](\boldsymbol{q})$ quantifies the difference
between the Hessian for structure $s$ obtained from the model
($\boldsymbol{K}_{{\rm eff},\boldsymbol{q}}^{s}$) and its
first-principles counterpart ($\boldsymbol{K}_{\boldsymbol{q}}^{s}$);
we define it as
\begin{equation}
{\mathcal D}^{s}[{\mathcal P}](\boldsymbol{q}) = \sum_{j} \left\|
\boldsymbol{K}_{{\rm eff},\boldsymbol{q}}^{s}
\hat{v}^{s}_{\boldsymbol{q}j} -
\lambda^{s}_{\boldsymbol{q}j}\hat{v}^{s}_{\boldsymbol{q}j}
\right\|^{2} \, ,
\end{equation}
where $\hat{v}^{s}_{\boldsymbol{q}j}$ and
$\lambda^{s}_{\boldsymbol{q}j}$ stand, respectively, for the
eigenvectors and eigenvalues of the first-principles Hessian
$\boldsymbol{K}_{\boldsymbol{q}}^{s}$. This strategy to compare the
Hessian matrices allowed us to achieve meaningful parameters in a
reliable and robust way; in contrast, we found that simpler schemes,
based on a direct comparison of eigenvalues or eigenvectors, lead to
difficult optimization problems that present many spurious local
minima of the goal function.

The above functions can be combined to run optimizations targeting
simultaneously at different properties. However, it is not clear {\sl
  a priori} how to weight the different goal functions in order to
construct a single ${\mathcal GF}$ that renders a well-posed
optimization problem. Hence, we adopted the following alternative
approach, which we used to generate most of the results presented in
Section~\ref{sec:application}. We start the parameter optimization by
minimizing one of the goal functions, ${\mathcal GF}_{1}$. Then a
second goal function ${\mathcal GF}_{2}$ is minimized, with the
parameters subject to the constraint that the result for ${\mathcal
  GF}_{1}$ must be preserved within a certain tolerance. In this way,
successive optimizations can be performed, with constraints involving
all previously-optimized goal functions, until we impose all the
necessary conditions. Naturally, the tolerances for the constraints
can be chosen so that the most critical properties are reproduced
better. Typically, in our work with PbTiO$_{3}$ and SrTiO$_{3}$ we
started by minimizing ${\mathcal GF}_{E}$, as we prioritize that our
models reproduce correctly the first-principles energies of the
structures in the training set. Then, the most usual sequence of
optimizations involved ${\mathcal GF}_{\nabla E}$, ${\mathcal GF}_{\rm
  hess}$ evaluated at the $\Gamma$ (i.e., $\boldsymbol{q} = 0$) point
of the lowest-energy structure(s), and finally ${\mathcal GF}_{\rm
  hess}$ evaluated at selected zone-boundary $q$-points of the
lowest-energy structure(s).

The optimization of ${\mathcal GF}_{\rm hess}$ was never prioritized
in the applications considered in this work; in fact, we found that,
when working with relatively simple (low-order) models as the ones
considered here, it is not realistic to aim at a very precise
description of the first-principles Hessians of structures that
deviate significantly from the RS. Nevertheless, we found that it was
often possible to adjust the low-lying eigenmodes at a reduced number
of $q$-points. Also, generally speaking, we found that considering
${\mathcal GF}_{\rm hess}$ was a good strategy to obtain
energy-bounded potentials, as such an optimization step helps to
impose the stability of the ground state structure.

\subsubsection{Further comments on the long-range
  interactions \label{sec:long-range-pars}}

As mentioned above, the atomic interactions in insulators can be
conveniently decomposed in short- and long-range parts. Further, at
the harmonic level we have a simple analytical expression for the
dipole-dipole coupling [Eq.~(\ref{eq:lr})] that depends on the RS
geometry, the dynamical charges $\boldsymbol{Z}^{*}_{i}$, and the
dielectric tensor $\boldsymbol{\epsilon}_{\infty}$. Conveniently,
these tensors, as well as the the decomposition of
$\boldsymbol{K}^{(2)}$ into $\boldsymbol{K}^{(2),sr}$ and
$\boldsymbol{K}^{(2),lr}$, are produced automatically by most DFPT
implementations; in particular, they are readily provided by {\sc
  Abinit}. [The typical DFPT scheme computes the {\em total}
  interatomic force constants. Then, in essence, it is assumed that
  the long-range part $\boldsymbol{K}^{(2),lr}$ is given by the
  dipole-dipole term in Eq.~(\ref{eq:lr}), and the short-range part is
  obtained as $\boldsymbol{K}^{(2),sr}$ = $\boldsymbol{K}^{(2)}$ $-$
  $\boldsymbol{K}^{(2),lr}$.] Alternatively, all the relevant
parameters controlling the dipole-dipole interactions can be obtained
by considering the response to finite electric fields.\cite{souza02}

As regards the anharmonic terms, we could continue to distinguish
between short- and long-range couplings. In essence, the anharmonic
long-range couplings in $E_{\rm p}$ would capture the changes in the
effective charges or dielectric constants that may be caused by the
atomic displacements and which affect the magnitude of the Coulombic
dipole-dipole interactions. As regards the strain-phonon couplings in
$E_{\rm sp}$, an additional effect comes from the change in the cell
shape and dimensions.

Our model potentials provide a framework to capture such effects by
considering appropriate high-order terms. Unfortunately, considering
such couplings would result in computationally-heavy atomistic
simulations. Indeed, as discussed in Section~\ref{sec:practical}, for
a practical implementation of the harmonic long-range interactions it
is convenient to precalculate, for the RS geometry and our specific
choice of simulation supercell, the dipole-dipole couplings by
performing the appropriate Ewald sums. Once the interaction
coefficients $\boldsymbol{K}^{(2),lr}$ are known, the corresponding
energy can be readily obtained during the course of the simulation;
yet, because such a term couples all the atoms in the supercell, its
calculation is by far the most time-consuming part of the energy
evaluation. In principle, one may proceed similarly with the
higher-order long-ranged terms. For example, consider
\begin{equation}
\boldsymbol{\Lambda}^{(1,2)} = \boldsymbol{\Lambda}^{(1,2),sr} +
\boldsymbol{\Lambda}^{(1,2),lr} \, ,
\end{equation}
which is the leading strain-phonon coupling for materials like
PbTiO$_3$ and SrTiO$_3$. In this case, we can model
$\boldsymbol{\Lambda}^{(1,2),lr}$ by considering the dependence of
$\boldsymbol{K}^{(2),lr}$ [Eq.~(\ref{eq:lr})] on a strain
$\boldsymbol{\eta}$ to linear order. (To do this, one could proceed by
introducing the strain dependence of the effective charges, dielectric
tensor, and interatomic distances in Eq.~(\ref{eq:lr}), and then
Taylor expand with respect to $\boldsymbol{\eta}$.) The corresponding
coefficients could be precomputed for the RS geometry and particular
simulation supercell, which would permit an easy (but still
computationally costly) evaluation of such an energy contribution
during the course of the simulation.

In our work with PbTiO$_3$ and SrTiO$_3$, wanting to obtain models
that allow for fast simulations, we did not treat explicitly the
anharmonic corrections to the long-range dipole-dipole
interactions. Yet, we captured the effects on the properties of
interest (e.g., the energy, equilibrium structure, and Hessian of
low-lying phases) in the short-range anharmonic couplings. Whenever
the anharmonic couplings are determined by the fitting procedure
outlined in Section~\ref{sec:fit-pars}, this can be done in the most
natural way. We simply assume that
$\widetilde{\boldsymbol{K}}^{(n),lr} =
\widetilde{\boldsymbol{\Lambda}}^{(n,m),lr} = 0$, and fit the
anharmonic terms $\widetilde{\boldsymbol{K}}^{(n),sr}$ and
$\widetilde{\boldsymbol{\Lambda}}^{(n,m),sr}$ to reproduce
first-principles information about the structures in our training set,
thus capturing effectively the consequences of possible
anharmonicities in the long-range couplings.

Additionally, we also computed the strain-phonon couplings directly,
without performing any fit, by proceeding in the following way. We
considered the full interatomic constants for the RS and strained
configurations, and assumed that the following approximate version of
Eq.~(\ref{eq:finitediff-lambda})
\begin{equation}
\begin{split}
&K^{(2),sr}_{i\alpha j\beta}\Bigr|_{\delta\boldsymbol{\eta}} +
  K^{(2),lr}_{i\alpha j\beta} \Bigr|_{\delta\boldsymbol{\eta}}\\
& \;\;\; \approx K^{(2),sr}_{i\alpha j\beta} +
K^{(2),lr}_{i\alpha j\beta} + \sum_{a} \Lambda^{(1,2),sr}_{a i\alpha
  j\beta} \delta\eta_{a}
\end{split}
\label{eq:finitediff-lambda-shortrange}
\end{equation}
holds within a certain spatial range (i.e., for a maximum separation
of atoms $i$ and $j$). Then, we demanded that the short-range part of
$\boldsymbol{\Lambda}^{(1,2)}$ capture strain-induced changes in both
$\boldsymbol{K}^{(2),sr}$ and $\boldsymbol{K}^{(2),lr}$. It must be
noted that, because of the spatial truncation, the
$\boldsymbol{\Lambda}^{(1,2),sr}$ thus calculated will in general
break translational invariance. To remedy this, we added to
$\boldsymbol{\Lambda}^{(1,2),sr}$ a correction
$\Delta\boldsymbol{\Lambda}^{(1,2),sr}$ that was determined by
demanding that our model reproduce exactly the Hessian of the strained
configurations at the $\Gamma$ point. In this way, by imposing a
correct description of the acoustic modes, we restore the
ASR. Further, this procedure also guarantees that the effect of strain
on the $\Gamma$ distortions, which are critical for the investigation
of ferroic perovskites like ours, is captured by our models. As shown
in Section~\ref{sec:pto}, this approximation leads to a very precise
description of the strain effects on the force-constant bands in the
case of PbTiO$_{3}$.

\section{Examples of application\label{sec:application}}

Now we describe the model potentials for ferroic perovskites
PbTiO$_{3}$ (PTO) and SrTiO$_{3}$ (STO) that we constructed following
the above scheme. These materials are representative of the large
family of compounds undergoing structural phase transitions driven by
soft phonon modes. The lattice dynamical properties of such systems
are strongly anharmonic, and the description of their transitions
requires the use of high-order potentials. Further, in the case of
these perovskite oxides the relevant energy scale for the soft mode
instabilities is relatively small, of about 50~meV {\sl per} formula
unit (f.u.) or less. Hence, achieving a good description of such
compounds constitutes a challenge for first-principles theory and,
naturally, for our model-potential approach.

Additionally, PTO and STO present peculiarities that make them
especially interesting in the present context. At temperature $T_{\rm
  C}$~=~760~K, PTO undergoes a transition between the high-$T$
paraelectric structure (i.e., the ideal cubic perovskite prototype,
with space group $Pm\bar{3}m$, that we take as our RS) and its low-$T$
ferroelectric (FE) phase (with tetragonal space group
$P4mm$).\cite{lines-book1977,haun87} The structural distortion that
appears at low temperatures has a polar character, and it essentially
involves a displacement of the Ti and Pb cation sublattices against
the O$_6$-octahedron network, as sketched in
Fig.~\ref{fig:instabilities}(a). Note that this corresponds to the
condensation of a soft mode at the zone center (at the $\Gamma$ point)
of the BZ of the RS. This transition has a significant first-order
character that previous theoretical work has linked with the
accompanying deformation of the cell;\cite{waghmare97} further,
first-principles theory predicts that cell strains are critical to
determine the symmetry of the ground state of PTO.\cite{kingsmith94}
Hence, to model this compound we have to deal with both the FE
instability responsible for the transformation at $T_{\rm C}$ {\em
  and} the strain-phonon couplings that have a strong impact in the
occurring equilibrium phases and the features of the FE transition.

SrTiO$_3$ too undergoes a single phase transition, as it transforms at
105~K from the high-$T$ cubic perovskite phase to a low-$T$ structure
of tetragonal ($I4/mcm$) symmetry.\cite{lines-book1977} The structural
distortion occurring in the low-$T$ phase involves concerted rotations
of the O$_6$ octahedra about the tetragonal axis, with the peculiarity
that O$_6$ groups that are first neighbors along $z$ rotate in
antiphase. Such a pattern is denoted $a^{0}a^{0}c^{-}$ in the
well-known notation introduced by Glazer,\cite{glazer75} and
corresponds to a so-called antiferrodistortive (AFD) mode associated
with the $R$ point of the BZ of the RS [$\boldsymbol{q}_{R} =
  \pi/a_{0} (1,1,1)$, where $a_{0}$ is the lattice constant of the RS
  cubic unit cell]; the corresponding atomic displacements are
sketched in Fig.~\ref{fig:instabilities}(c). Additionally, STO is
close to presenting a FE instability; in fact, this compound is
experimentally believed to be a {\em quantum-paraelectric}, i.e., a
material whose ferroelectricity is suppressed by quantum fluctuations
(i.e., the wave-like character) of its constituting
atoms.\cite{muller79,viana94,zhong96} Further, previous
first-principles work has shown that the FE and AFD soft modes compete
in STO,\cite{zhong95b} complicating even more the description of the
behavior of the material at low temperatures. Hence, STO offered us
the possibility of testing our approach in cases in which several
structural instabilities are relevant and their interaction must be
considered in detail.

We first describe our work with PTO, which turned out to present all
the challenging features that we had anticipated (i.e., the very
critical strain-phonon couplings) and additional ones that we were not
expecting (i.e., a very significant competition between FE and AFD
modes). Hence, we discuss the case of PTO in detail, giving
illustrative examples of how our models can be extended when it is
necessary to do so. In contrast, it was relatively easy to obtain a
sound model for STO. Hence, in that case we will present a very
minimal approach to the construction of an effective potential. In
both cases, we will describe the $T$-driven transitions obtained when
solving our models by means of Monte Carlo simulations (in which, as
usually done, we treated atoms as classical objects), showing that
they capture correctly the basic experimental behaviors. We will also
comment on the probable origin of the quantitative discrepancies
observed between our model predictions and experiment. Note that here
we will not elaborate much on the physics emerging from our models,
as such a discussion falls beyond of the scope of this paper.

\subsection{First-principles and Monte Carlo
  methods \label{sec:methods}} 

All first-principles calculations were done with the {\sc Abinit}
package,\cite{abinit09} and employed the local-density approximation
(LDA) to density functional theory.\cite{hohenberg64,kohn65} The ionic
cores were treated by using extended norm-conserving Teter
pseudopotentials,\cite{teter93} and the following electrons were
considered explicitly in the calculations: Pb's 5$d^{10}$, 6$s^{2}$,
and 6$p^{2}$; Sr's 4$s^{2}$, 4$p^{6}$, and 5$s^{2}$; Ti's 3$s^2$,
3$p^6$, 3$d^2$, and 4$s^2$; and O's 2$s^2$ and 2$p^4$. Electronic wave
functions were represented in a plane-wave basis truncated at
1500~eV. We used an $8\times 8\times 8$ $k$-point grid to compute
integrals in the Brillouin-zone of the 5-atom perovskite cell, and
equivalent meshes for other cells. In structural relaxations, atomic
positions were optimized until residual forces on atoms were below
10$^{-4}$~eV/{\AA}. The interatomic force constants, elastic
constants, Born charges, and dielectric tensor were calculated by
using the DFPT implementation in {\sc Abinit}. The
$\boldsymbol{K}_{\boldsymbol{q}}^{s}$ matrices, from which the
real-space interatomic constants are obtained, were computed for a
$2\times 2\times 2$ $q$-point mesh; in agreement with previous
studies,\cite{ghosez99} this was found to be sufficient to get
accurate results. The resulting cut-off radius for the short-range
interactions is therefore about 6.8~\AA\ for all of the presented
models.

Thermal averages of the quantities of interest were calculated by a
standard Metropolis Monte Carlo method.\cite{allen-book1989} The
Markov chain was constructed by sequentially considering movements of
(i.e., by {\em sweeping through}) all atoms in the simulation
box. After each sweep, a single attempt to modify each of the strain
components was made. Both the attempted displacements and strain
modifications were drawn from appropriate uniform distributions, whose
widths were varied by a simple linear controller with the goal of
attaining an acceptance rate of 50\% on average. In most cases, we
used an $8\times 8\times 8$ periodically-repeated simulation
supercell, and thermalized the material by running 20000 MC sweeps
starting from the RS (i.e., $\boldsymbol{u}_i = \boldsymbol{\eta} =
0$). The averages for the relevant structural distortions were then
calculated from 20000-40000 additional sweeps, and we checked
convergence by inspection of the corresponding histograms. At
temperatures in the vicinity of the phase transitions, this procedure
did not lead to converged results because of either slow
thermalization or finite-size effects. In such cases, we found it
necessary to run the calculations of up to 80000 sweeps in $10\times
10\times 10$ simulation supercells. For presentation purposes, all of
the computed average distortions were rotated so that the axes for the
FE polarization and AFD rotations lie along the [001] (resp. [111])
Cartesian direction for tetragonal (resp. rhombohedral) phases.

\subsection{PbTiO$_{3}$ \label{sec:pto}}

\subsubsection{Harmonic terms $E_{\rm har}(\{\boldsymbol{u}_{i}\})$ and
  $E_{\rm s}(\boldsymbol{\eta})$}

The first step in the construction of our model potential is the
computation of the harmonic energy terms, $E_{\rm har}$ and $E_{\rm
  s}$, for which we use the DFPT scheme\cite{gonze97,wu05} implemented
in {\sc Abinit}.\cite{abinit09} (As mentioned above, the third
harmonic term -- i.e., the strain-phonon coupling
$\boldsymbol{\Lambda}^{(1,1)}$ in $E_{\rm sp}$ -- is identically zero
in PTO and STO due to the cubic symmetry of the RS.)  Representative
results are given in Fig.~\ref{fig:phonons-pto}, which shows the bands
corresponding to the stiffness coefficients or force constants of the
cubic RS. (These are the eigenvalues $\lambda_{\boldsymbol{q}j}$ of
the Hessian matrices $\boldsymbol{K}_{{\rm eff},\boldsymbol{q}}$
introduced in Section~\ref{sec:fit-pars}.) Notably, we find that some
distortions have a negative stiffness, indicating that they are
structural instabilities of the RS. The leading instabilities are
pictorially represented in Fig.~\ref{fig:instabilities}: the FE soft
mode at the $\Gamma$ point [panel~(a)], the in-phase AFD mode at the
$M$ point [$\boldsymbol{q}_{M} = \pi/a_{0} (1,1,0)$, where $a_{0}$ is
  the lattice constant of the RS cubic unit cell; panel~(b)], and the
antiphase AFD mode at the $R$ point [panel~(c)]. As shown in
Fig.~\ref{fig:phonons-pto}, the energetics of all such instabilities
is captured exactly, at the harmonic level, by our model.

\begin{figure}
 \includegraphics[width=0.9\columnwidth]{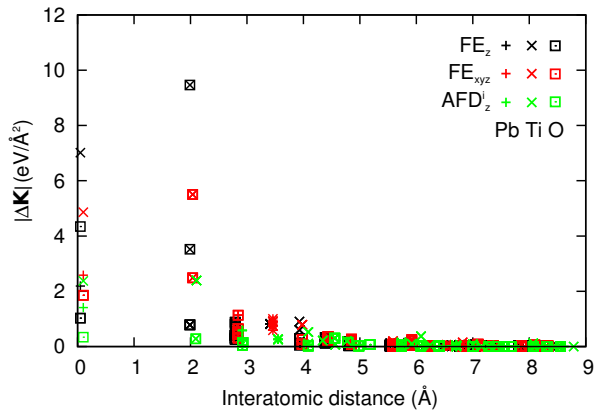}
\caption{(Color on line.) Deviation of the interatomic force constants
  calculated for selected PTO distorted structures from the RS
  results. Each point quantifies the difference between the 3$\times$3
  force-constant matrices, for a specific atom pair, computed for the
  reference and distorted structures. The results are shown as a
  function of interatomic distance; for clarity reasons, the
  interatomic distances are shifted slightly to reduce overlap. Note
  that, when two different atoms are involved in the pair, we overlap
  the corresponding symbols; thus, for example, crossed squares
  correspond to pairs involving Ti (cross) and O (square) atoms.}
\label{fig:local-anh}
\end{figure}

As a result of our DFPT calculations, we obtained an $E_{\rm har}$
term that includes all short-range interactions within a spatial range
slightly below 7~\AA. (For example, this includes couplings between Ti
pairs that are 3rd nearest neighbors.) Additionally, $E_{\rm har}$
includes the already-mentioned analytic form of the long-range
dipole-dipole couplings, which involves 5 symmetry-independent
parameters [i.e., 4~Born effective charges (which reduce to 3
  independent ones if the ASR in Eq.~(\ref{eq:ZASR}) is considered)
  and 1~dielectric constant that fully defines the diagonal and
  isotropic tensor]. As regards the harmonic elastic constants in
$E_{\rm s}$, the model incorporates the 3 symmetry-independent terms
that define the full elastic tensor for a crystal with cubic
$m\bar{3}m$ point symmetry.

\subsubsection{Fitting $E_{\rm anh}(\{\boldsymbol{u}_{i}\})$}

Next we tackled the construction of the anharmonic terms of the
potential. We first considered the case in which the cell is fixed to
be that of the RS, i.e., we assumed $\boldsymbol{\eta} = 0$ and
focused on $E_{\rm anh}$. As described above, we computed $E_{\rm
  anh}$ by fitting its parameters to a set of relevant
first-principles data. Naturally, we populated our training set with
information about the low-energy structures that can be accessed by
{\em condensing} the different instabilities of the RS. More
specifically, our list of low-symmetry phases contains FE structures
of tetragonal (FE$_z$) and rhombohedral (FE$_{xyz}$) symmetries, as
well as several AFD-distorted phases (AFD$_{z}^{a}$, AFD$_{xyz}^{a}$,
and AFD$_{z}^{i}$). [We use the notation of
  Fig.~\ref{fig:instabilities}, with the $xyz$ subscript denoting the
  simultaneous occurrence of a distortion type along/about the three
  Cartesian axes, and with the same amplitude for the three of them.]
In addition, we also considered a hybrid structure, FE$_{xyz}$ $+$
AFD$_{xyz}^{a}$, that has rhombohedral $R3c$ symmetry and in which
both FE and AFD patterns occur simultaneously. Note that we determined
such low-symmetry structures {\sl ab initio} by ($i$) distorting the
RS according to a specific soft-mode eigenvector and then ($ii$) using
this as the starting point of a structural relaxation that preserves
the symmetry of the initial configuration. In all cases we computed
the equilibrium structure, energy, and Hessian matrix. The relevant
structural parameters of the considered phases are given in
Table~\ref{tab:PTO_structs}, together with the energies relative to
the RS.

\begin{table*}
\caption{Structural parameters of the considered low-symmetry
  PbTiO$_3$ structures (see the text) calculated for a cubic cell with
  lattice constant $a$~=~3.880\AA. First-principles LDA results are
  presented along with those obtained from the model discussed in the
  text. We show the $\Gamma$-point displacements corresponding to the
  polar distortion, $u^{\Gamma}_{\kappa\alpha}$, given in Angstrom; we
  chose them so that there is no rigid shift of the whole structure
  (i.e., $u_{{\rm Pb}z} + u_{{\rm Ti}z} + 2u_{{\rm O1}z} + u_{{\rm
      O3}z} = 0$). For $P4mm$, all the displacements are along the $z$
  direction and we have $u_{{\rm O1}z}^{\Gamma} = u_{{\rm
      O2}z}^{\Gamma}$. For $R3m$ and $R3c$, we have $u_{\kappa
    x}^{\Gamma}$ = $u_{\kappa y}^{\Gamma}$ = $u_{\kappa z}^{\Gamma}$
  for the Pb and Ti atoms, as well as $u^{\Gamma}_{{\rm O3}z}$ =
  $u^{\Gamma}_{{\rm O1}x}$ = $u^{\Gamma}_{{\rm O2}y}$ and
  $u^{\Gamma}_{{\rm O1}z}$ = $u^{\Gamma}_{{\rm O1}y}$ =
  $u^{\Gamma}_{{\rm O2}x}$ = $u^{\Gamma}_{{\rm O2}z}$ =
  $u^{\Gamma}_{{\rm O3}x}$ = $u^{\Gamma}_{{\rm O3}y}$. The amplitude
  of the AFD modes is quantified by the corresponding O$_{6}$ rotation
  angle given in degrees (in the rhombohedral cases, we have
  equal-magnitude rotations about the three Cartesian axes; we give
  the rotation angle for one axis.). Energies are given in meV/f.u.,
  taking the result for the RS as the zero of energy.}
\begin{tabular}{ccdddddd}
\hline\hline
Structure & Method & \multicolumn{1}{c}{$u_{{\rm Pb}z}^{\Gamma}$} 
                   & \multicolumn{1}{c}{$u_{{\rm Ti}z}^{\Gamma}$} 
                   & \multicolumn{1}{c}{$u_{{\rm O1}z}^{\Gamma}$} 
                   & \multicolumn{1}{c}{$u_{{\rm O3}z}^{\Gamma}$} 
                   & \multicolumn{1}{c}{O$_{6}$ rot.} 
                   & \multicolumn{1}{c}{Energy} \\\hline
\multirow{2}{*}{FE$_z$ ($P4mm$)}     & LDA    &  0.179     & 0.072      & -0.104     &  -0.043        & -       & -23.7 \\
&  model   &  0.180     & 0.073      & -0.105  & -0.043        & -       & -24.8 \\
\multirow{2}{*}{FE$_{xyz}$ ($R3m$)}    & LDA    &  0.104     & 0.048      & -0.063     & -0.027       & -       & -26.6 \\
        & model   &  0.105     & 0.049      & -0.063     & -0.028
& -       & -28.3 \\
\multirow{2}{*}{AFD$_{z}^{a}$ ($I4/mcm$)}     & LDA    &  -       & -        & -       & -        & 5.4       & -9.4 \\
       & model   &  -       & -        & -       & -        &
5.9       & -11.7 \\
\multirow{2}{*}{AFD$_{xyz}^{a}$ ($R\bar{3}c$)}   & LDA    &  -       & -        &  -       & -        & 3.4       & -11.2 \\
   & model   & -       & -        & -       & -      & 3.4       & -11.7 \\
\multirow{2}{*}{AFD$_{z}^{i}$ ($P4/mbm$)}     & LDA    &  -       & -        &  -       &  -       & 3.9       & -2.7 \\
      & model   &  -       & -        & -       & -        & 4.3       & -3.3 \\
\multirow{2}{*}{FE$_{xyz}$+AFD$_{xyz}^{a}$ ($R3c$)}   & LDA    &  0.096     & 0.047      &  -0.058     & -0.026      & 2.8       &
-29.5 \\
      & model   &  0.098     & 0.047      &  -0.060     & -0.026       & 2.1       & -29.5 \\\hline\hline
\end{tabular}
\label{tab:PTO_structs}
\end{table*}

In order to fit $E_{\rm anh}$, we worked with the
displacement-difference representation and
$\widetilde{\boldsymbol{K}}^{(n)}$ parameters (with $n > 2$) of
Eq.~(\ref{eq:Eu-differences}). We restricted ourselves to models that
include only pairwise interactions and extend up to 4th order in the
Taylor series. These approximations define the minimal model needed to
capture structural phase transitions like the ones we want to
describe, and are analogous to the ones adopted in most of the
previous theoretical works that we are aware of. (One of the few
exceptions is the inclusion of high-order terms for the local polar
modes considered in Ref.~\onlinecite{waghmare97}.) In our case, we
maintained such approximations in order to keep our models relatively
simple and computationally efficient, as well as to test the actual
ability of such an elementary potential to reproduce the
first-principles data in a quantitative way.

\begin{figure}[t!]
 \includegraphics{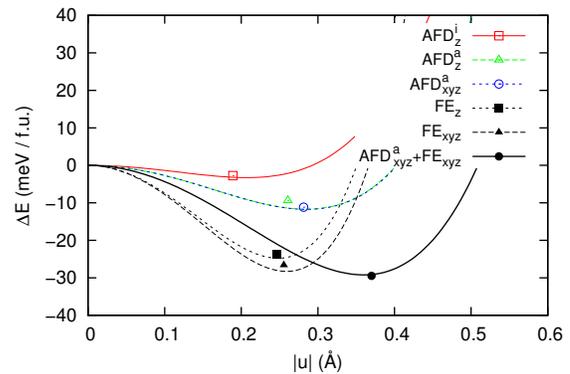}
 \caption{(Color on line.) Potential-energy wells connecting the RS of
   PTO with the low-symmetry phases defined in the text. The results
   obtained from our model potential are shown with lines, and the
   points indicate the first-principles results for the energy minima
   or saddles. All the states shown preserve the cubic cell of the RS
   ($\boldsymbol{\eta} = 0$). The amplitudes $|u|$ in Angstrom
   correspond to collective distortions involving several atoms. The
   AFD$_{z}^{a}$ and AFD$_{xyz}^{a}$ curves are essentially on top of
   each other and cannot be distinguished.}
 \label{fig:PTO_fit}
\end{figure}

As regards the spatial extent of the anharmonic couplings, most of the
previous works on phenomenological models and effective Hamiltonians
adopt what is sometimes called the {\em on-site anharmonicity}
approximation, which implies that the non-harmonic couplings are taken
to be strictly confined in space and contribute only to the
self-energy of the atoms or local
modes.\cite{blinc-book1974,lines-book1977,zhong94a,zhong95a,waghmare97}
Interestingly, our first-principles results give us a direct way to
test whether such an approximation is justified.
Figure~\ref{fig:local-anh} shows the difference between the harmonic
interatomic couplings computed for the RS (which are given by
$\boldsymbol{K}^{(2)}$ directly) and those corresponding to the
several distorted states of PTO that maintain the cubic cell (which
are described by $\boldsymbol{K}^{(2)}$ plus a distortion-dependent
correction involving $\widetilde{\boldsymbol{K}}^{(n)}$ with $n >
2$). From these results, it is apparent that the distortion-induced
changes decay very rapidly with the interatomic distance, indicating
that the anharmonic corrections have a limited spatial range; similar
calculations for other distorted configurations confirmed this
conclusion. Hence, our model for PTO included only anharmonic
$\widetilde{\boldsymbol{K}}^{(n)}$ couplings between neighboring atom
pairs (i.e., each Pb atom is coupled with its 12 neighboring oxygens,
and each Ti atom with the 6 oxygens in the surrounding O$_{6}$ group),
which results in couplings extending up to about 3~\AA. Note that this
approximation is essentially equivalent to the on-site-anharmonicity
assumption of the effective-Hamiltonian method, but adapted to our
displacement-difference representation. Together with the other
truncations mentioned above (pairwise interactions, 4th-order Taylor
series), this local-anharmonicity approximation results in the 15 SATs
listed in Table~\ref{tab:SATs}.

\begin{figure*}[t!]
\begin{tabular}{ccc}
 (a) cubic ($Pm{\bar 3}m$) & (b) FE$_z$ ($P4mm$) & (c) FE$_{xyz}$ ($R3m$)
 \\ \includegraphics[width=.31\textwidth]{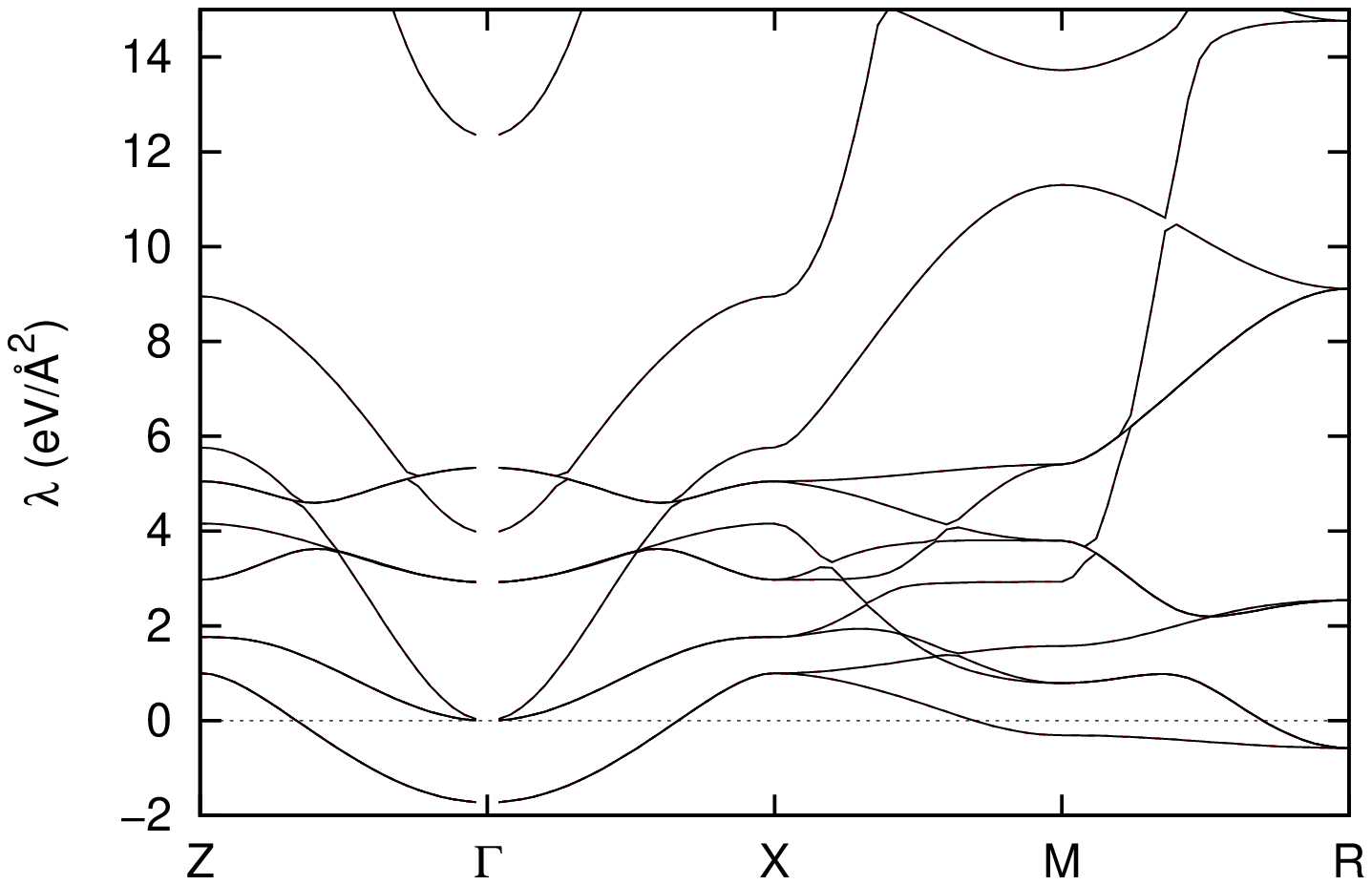} 
 &
 \includegraphics[width=.31\textwidth]{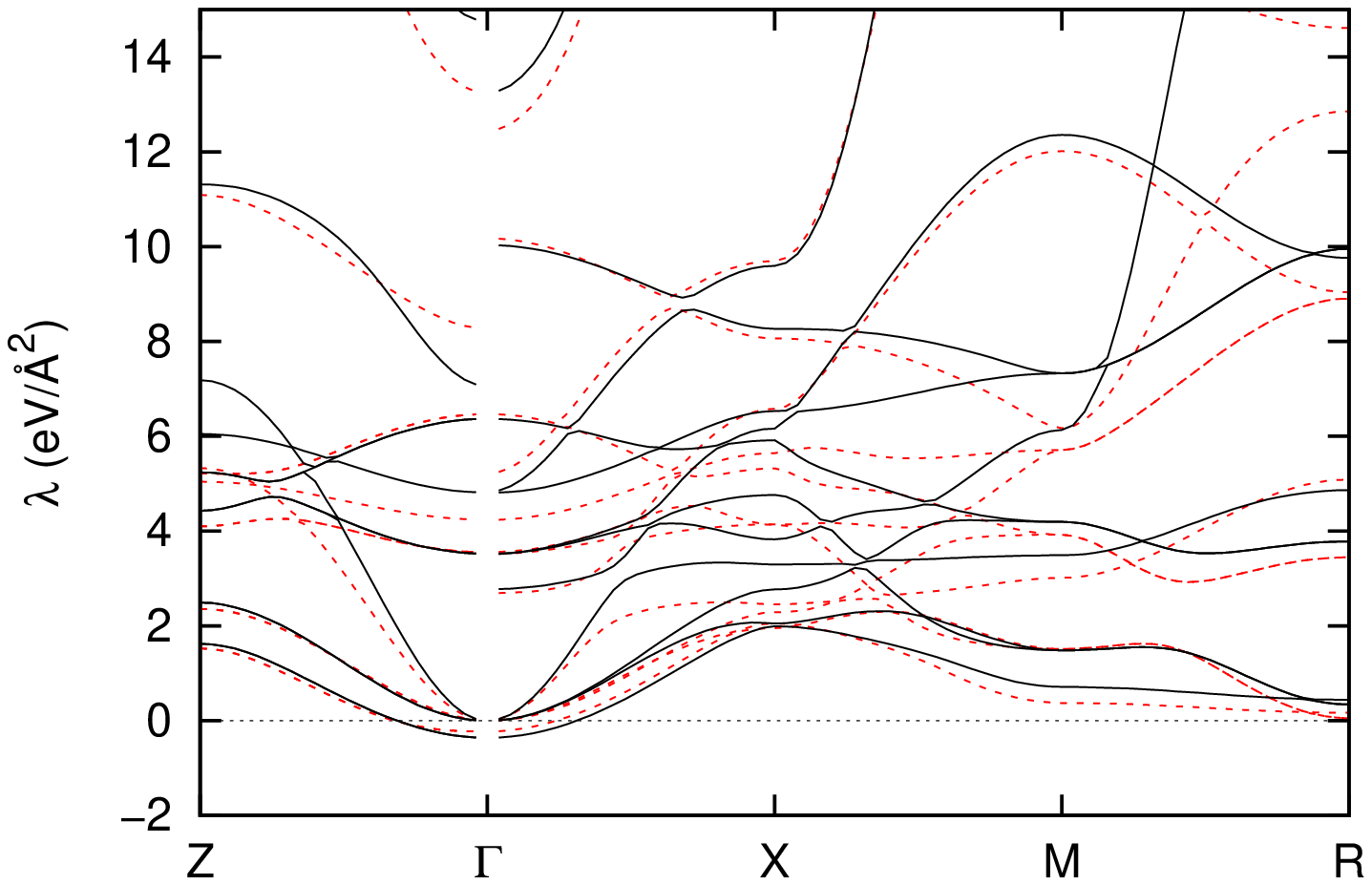}
 &
 \includegraphics[width=.31\textwidth]{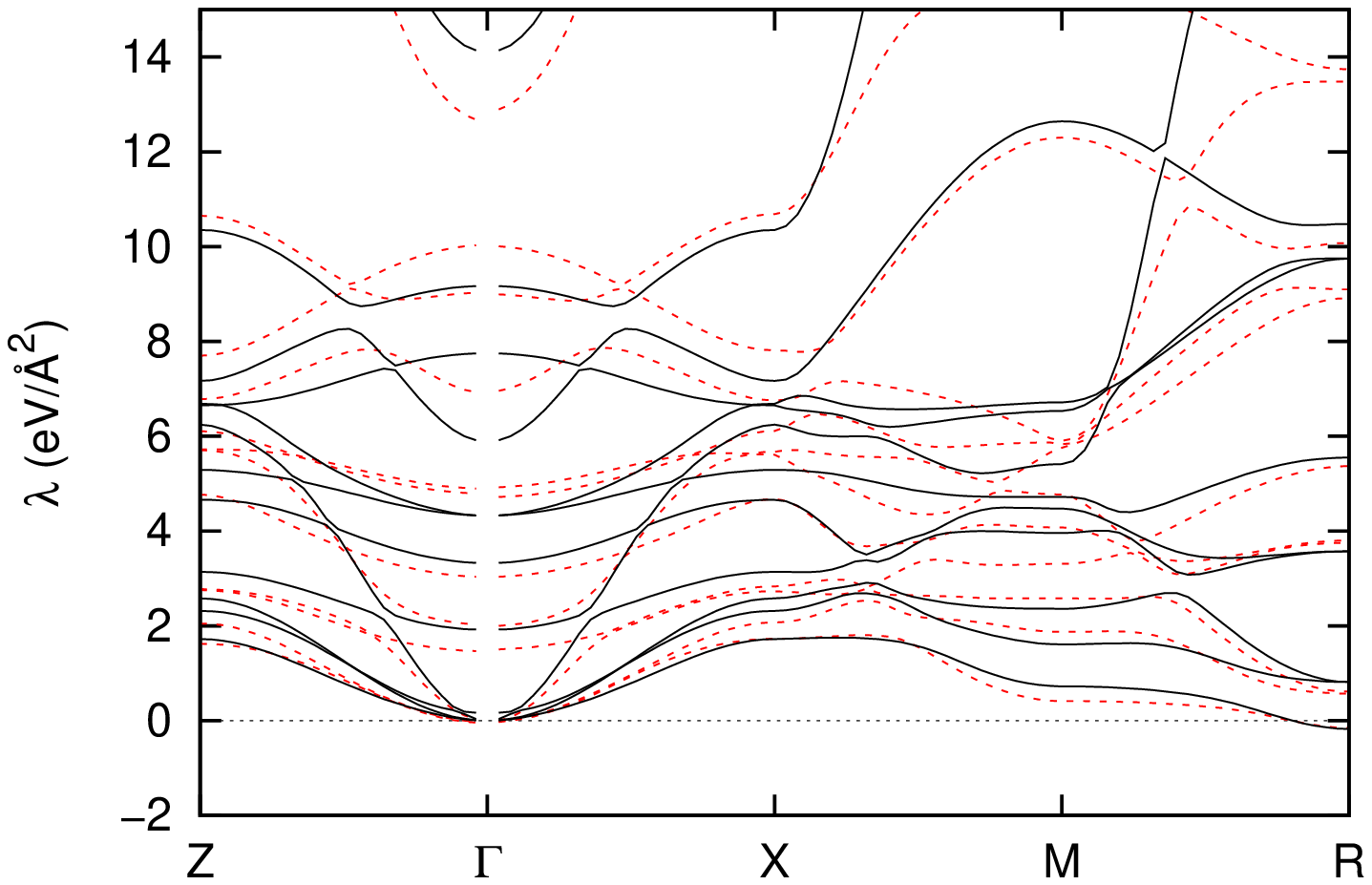}
 \\
\end{tabular}
\caption{(Color on line.) Force-constant bands corresponding to three
  different PTO structures, all maintaining the cubic RS cell. Black
  solid lines show the results of our model potential, and red dashed
  lines the first-principles results.}
\label{fig:bands-PTO-dist}
\end{figure*}

Using the model and training set described above, we fitted the 15
anharmonic parameters of Table~\ref{tab:SATs} by successive
optimization of the $\mathcal{GF}_{E}$, $\mathcal{GF}_{\nabla E}$, and
$\mathcal{GF}_{\rm hess}$ goal functions, following the procedure
outlined in Section~\ref{sec:fit-pars}. In $\mathcal{GF}_{\rm hess}$
we considered the Hessian matrices of distorted configurations,
including modes corresponding to the $\Gamma$ and, in some cases, $R$
points of the BZ of the RS. For each $q$-point, we considered only the
6 lowest-lying optical eigenmodes (i.e., we did not fit to the full
spectrum). As evidenced by Table~\ref{tab:PTO_structs} and
Fig.~\ref{fig:PTO_fit}, the model thus constructed describes with good
accuracy our first-principles results for the equilibrium structures
and energies of the relevant $\boldsymbol{\eta} = 0$
configurations. Additionally, Fig.~\ref{fig:bands-PTO-dist} shows the
results that our model gives for the force-constant bands of two
distorted structures; as expected, the low-lying Hessian eigenmodes
are reasonably well reproduced, and the inaccuracies grow as we move
up in energy.

\begin{figure}[b!]
\includegraphics[width=.9\columnwidth]{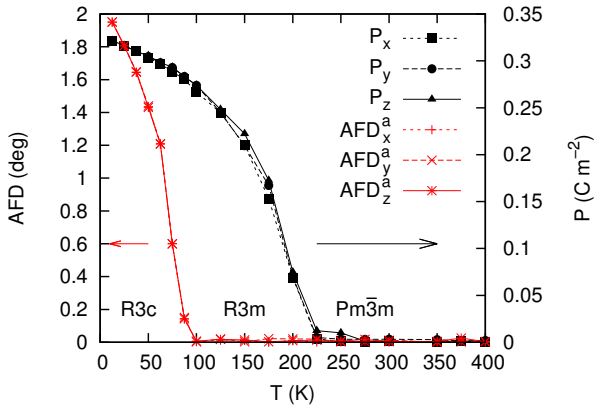} 
\caption{(Color on line.) Temperature-dependent polarization and
  AFD$^{a}$ order parameters of PTO as obtained from MC simulations of
  our model under the $\boldsymbol{\eta} = 0$ condition.}
\label{fig:PTO_LDA_fixed_tscan}
\end{figure}

To test our model for a fixed-cell version of PTO, we ran MC
simulations and computed the evolution of the equilibrium structure as
a function of temperature. Figure~\ref{fig:PTO_LDA_fixed_tscan} shows
our basic results, which reveal a sequence of two phase transitions:
At $T\approx 200$~K the material develops a spontaneous polarization,
which manifests itself in a non-zero value of the dipole moments
averaged over all cells in the simulation box. Such a transition
drives the system from its high-$T$ cubic ($Pm\bar{3}m$) phase to a
rhombohedral ($R3m$) one; the spontaneous polarization is parallel to
the rhombohedral axis, which lies along the [1,1,1] Cartesian
direction. Such a $R3m$ structure is usually thought to be the ground
state of PTO subject to the $\boldsymbol{\eta} = 0$
condition.\cite{kingsmith94,waghmare97} However, our MC simulations
rendered a second transition, at $T\approx 100$~K, in which an AFD
mode freezes in. More precisely, at low temperatures we observe the
occurrence of a distortion involving antiphase rotations of the
O$_{6}$ groups about all three Cartesian axes, which we denoted by
AFD$_{xyz}^{a}$ in the description above. The spontaneous polarization
remains essentially unaltered upon the condensation of this AFD mode,
and the new phase presents the $R3c$ rhombohedral space
group. Remarkably, this structure was not part of the initial training
set that we used to fit the parameters in $E_{\rm anh}$; indeed, we
discovered it by running MC simulations with our initial model
potentials for PTO, which led us to better characterize it from first
principles and eventually include it in the training set. This clearly
illustrates the usefulness of our model-potential approach to discover
new phenomena. [The force-constant bands of
  Fig.~\ref{fig:bands-PTO-dist}(c) already indicate that the $R3m$
  structure cannot be the ground state of our fixed-cell version of
  PTO. Note that the bands for the $R3m$ structure present a negative
  stiffness for some $R$-point modes, which correspond exactly to the
  low-$T$ AFD instability observed in the MC runs.]

\begin{table*}[t!]
  \caption{Results for the tetragonal ground-state structure of PTO
    with relaxed cell parameters. Lattice vectors and atomic
    displacements are given in Angstroms. The atomic displacements are
    as described in the caption of
    Table~\protect\ref{tab:PTO_structs}. We show the first-principles
    results (first row) followed by the results obtained from models
    with different descriptions of the strain-phonon coupling terms
    (see the text). Energies are given in meV/f.u., and we take
    $E_{\rm RS}$ as the zero of energy.}
  \begin{tabular}{ldddddddd}
\hline\hline
  Method & \multicolumn{1}{c}{a} 
         & \multicolumn{1}{c}{c} 
         & \multicolumn{1}{c}{c/a}  
         & \multicolumn{1}{c}{$u_{{\rm Pb}z}$} 
         & \multicolumn{1}{c}{$u_{{\rm Ti}z}$} 
         & \multicolumn{1}{c}{$u_{{\rm O1}z}$} 
         & \multicolumn{1}{c}{$u_{{\rm O3}z}$} 
         & \multicolumn{1}{c}{Energy} \\\hline
  LDA           & 3.864   & 3.974   & 1.029 & 0.230  & 0.106  & -0.133   & -0.071   & -37.7        \\
  model $L^{0}$& 3.908   & 3.987   & 1.020 & 0.200  & 0.103  & -0.122   & -0.060   & -34.5        \\
  model $L^{I}$& 3.863   & 3.968   & 1.027 & 0.220  & 0.099  &  -0.128  & -0.063   & -39.9        \\
  model $L^{II}$& 3.861   & 3.978   & 1.030 & 0.227  & 0.102  & -0.132 & -0.066    & -43.1   
   \\
  model $L^{III}$ & 3.856   & 3.968   & 1.029 & 0.221  & 0.098 & -0.128   & -0.062  & -39.9        \\
\hline\hline
  \end{tabular}
  \label{tab:PTO_tetragonal_relaxed}
\end{table*}

\subsubsection{The strain-phonon term $E_{\rm
    sp}(\{\boldsymbol{u}_{i}\},\boldsymbol{\eta})$}

We began by considering in $E_{\rm
  sp}(\{\boldsymbol{u}_{i}\},\boldsymbol{\eta})$ some of the
lowest-order terms that are not zero by symmetry, i.e., those
corresponding to the coefficients $\boldsymbol{\Lambda}^{(1,2)}$ or,
equivalently, $\widetilde{\boldsymbol{\Lambda}}^{(1,2)}$. This
constitutes the minimal approximation that captures the strain-phonon
couplings leading to physically relevant phenomena in ferroelectric
perovskites (e.g., piezoelectricity and the elastic effects associated
with the structural transitions), and is analogous to the one adopted
in the effective-Hamiltonian
literature.\cite{zhong94a,zhong95a,waghmare97}

\begin{figure*}
\begin{tabular}{ccc}
 (a) model $L^{0}$ & (b) model $L^{I}$ & (c) model $L^{III}$ \\
 \includegraphics[width=.31\textwidth]{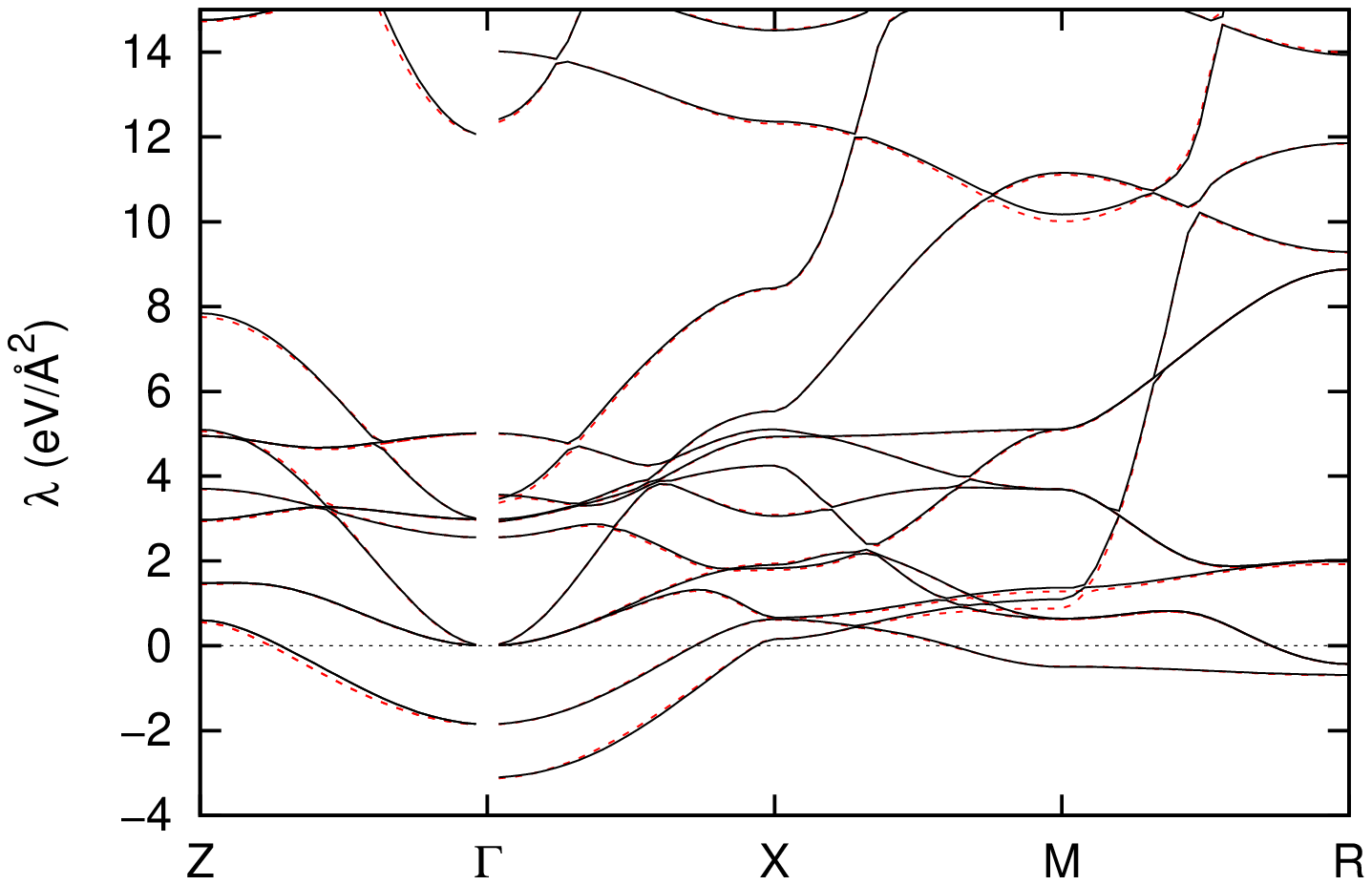} &
 \includegraphics[width=.31\textwidth]{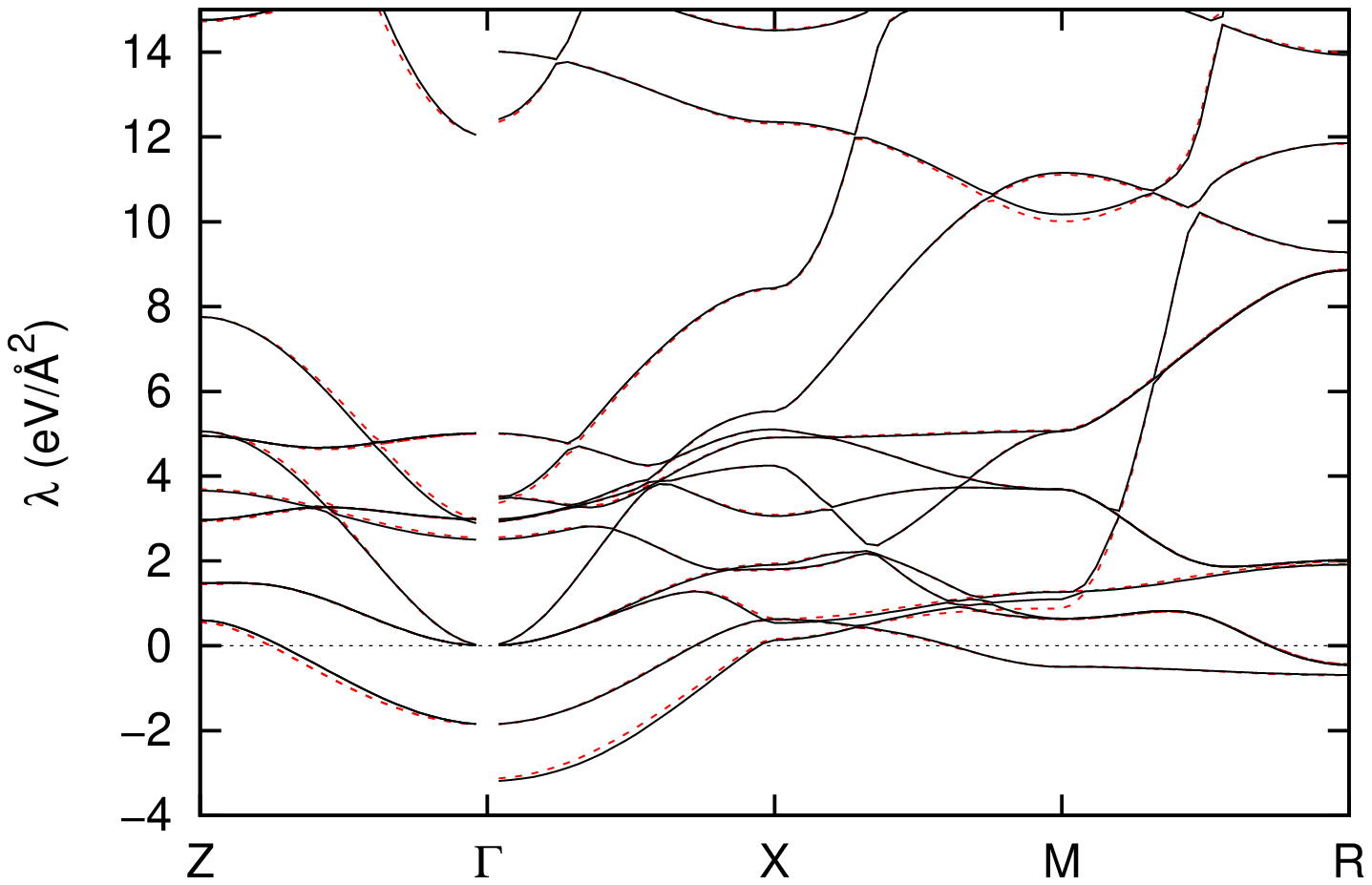} &
 \includegraphics[width=.31\textwidth]{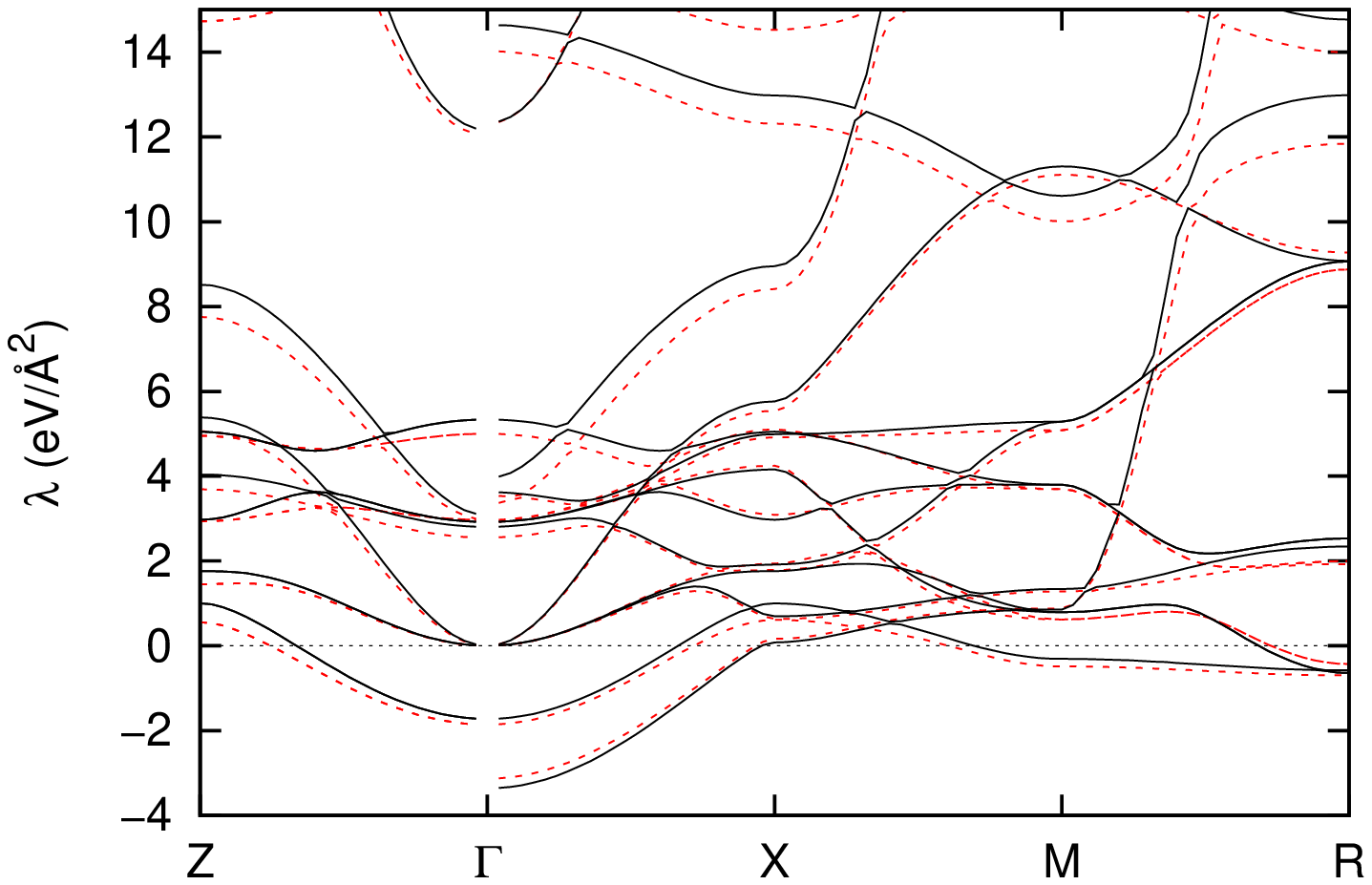} \\ 
\end{tabular}
\caption{(Color on line.) Force-constant bands corresponding to PTO
  structures with $\boldsymbol{u}_{i} = 0$ and subject to an uniaxial
  strain of 2\% (stretching along $z$), as obtained from three models
  that include different strain-displacement couplings (see the
  text). Black solid lines are the model results, and red dashed lines
  depict the bands obtained from first principles. Note the
  non-analytic behavior of the spectrum when approaching $\Gamma$
  from the $[q_{x},0,0]$ or $[0,0,q_{z}]$ directions.}
\label{fig:bands-PTO-strain}
\end{figure*}

As regards the spatial extent of the $\boldsymbol{\Lambda}^{(1,2)}$
interatomic couplings, the effective-Hamiltonian works have
traditionally adopted an on-site approximation that is analogous to
the one used for the anharmonic terms in $E_{\rm
  anh}$;\cite{zhong94a,zhong95a,waghmare97} consequently, only
one-body interactions are typically considered. Further, strain
effects on the long-range dipole-dipole interactions have never been
treated in the literature, to the best of our knowledge.

In our case, we went beyond such approximations by computing
$\boldsymbol{\Lambda}^{(1,2)}$ {\sl via} the approach described by
Eqs.~(\ref{eq:finitediff-lambda}) and
(\ref{eq:finitediff-lambda-shortrange}), using strains of $\pm$2\% for
the finite-difference calculations. (The spatial extent of the
$\boldsymbol{\Lambda}^{(1,2)}$ interactions thus computed is
essentially identical to that of the $\boldsymbol{K}^{(2)}$ terms.)
The model constructed in this way, which we call $L^{0}$, captures
very accurately the strain dependence of the force-constant bands, as
can be appreciated in Fig.~\ref{fig:bands-PTO-strain}(a); note that
such a good agreement throughout the BZ validates our approximate
method to treat the effect of strain on the long-range interactions
between dipoles, which was described in
Section~\ref{sec:long-range-pars}. Moreover, this $L^{0}$ model also
renders the correct low-$T$ structure for the real (unconstrained)
PTO: Indeed, the experimental ground state of bulk PTO at ambient
pressure is tetragonal ($P4mm$ space group), as opposed to the
rhombohedral ($R3c$) solution that we predict when imposing the
$\boldsymbol{\eta} = 0$ condition. Remarkably, the strain-phonon
couplings calculated with our finite-difference scheme capture such an
effect, even though they were not explicitly fitted to do so. On the
other hand, the predictions provided by this model do not reach the
quantitative accuracy of the results obtained in the fixed-cell
case. More precisely, Table~\ref{tab:PTO_tetragonal_relaxed} shows
significant differences between the first-principles results (labeled
``LDA'') and the predictions of the $L^{0}$ model for the structure of
the tetragonal ground state, especially as regards the aspect ratio
($c/a$) of the unit cell and the participation of the Pb atoms in the
ferroelectric distortion.

Wanting to increase the model's accuracy, we decided to improve the
description of the strain-phonon couplings by adding the SAT
represented by $({\rm Pb}_{x}-{\rm O2}_{x})^{2}\eta_{1}$, where we use
the compact notation introduced above (see Fig.~\ref{fig:SATs} and
Table~\ref{tab:etaSATs}). Note that the resulting model, which we
label $L^{I}$, combines $\Lambda$-like terms, whose values are fixed
to those of the $L^{0}$ potential, with one $\widetilde{\Lambda}$-like
free adjustable parameter. Such a parameter was fitted to better
reproduce the ground state structure; as shown in
Table~\ref{tab:PTO_tetragonal_relaxed}, this led to a significant
improvement over the $L^{0}$ result. Further improvement of the $c/a$
value can be achieved by additionally introducing the higher-order SAT
represented by $({\rm Pb}_{x}-{\rm O2}_{x})^{2}\eta_{1}^{2}$ (model
$L^{II}$), at the expense of worsening the agreement for other
structural parameters and energies.

Let us mention here another model-construction experiment that we
made. Noting the importance of the strain-phonon couplings in PTO, one
may wonder which are the interaction terms responsible for the main
effects. By inspecting the $\boldsymbol{\Lambda}^{(1,2)}$ parameters
computed directly from first principles, it is easy to identify the
two most prominent ones, which involve Pb--O and Ti--O
nearest-neighboring pairs. More specifically, the key couplings are
captured by the $\widetilde{\Lambda}$-like parameters number~2 and
number~13 from Table~\ref{tab:etaSATs}. Hence, we considered a model
that includes only these two strain-phonon couplings ($L^{III}$);
interestingly, as shown in Table~\ref{tab:PTO_tetragonal_relaxed},
such a simple potential is able to render good results for the
structure and energy of PTO's ground state.

\begin{figure}[t!]
\includegraphics[width=.9\columnwidth]{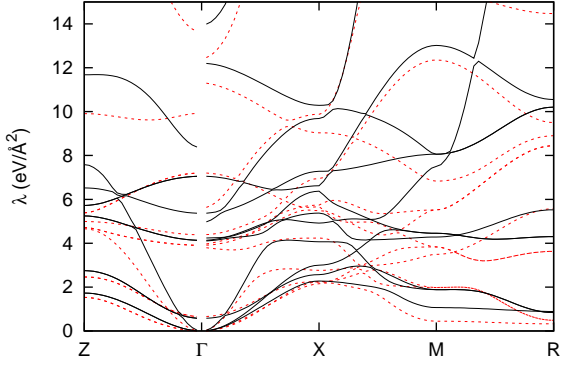}
\caption{(Color on line.) Force-constant bands for the ground state of
  PTO. Black solid lines depict the results from model $L^{I}$, and
  red dashed lines show the results obtained from first principles.}
\label{fig:bands-PTO-gs}
\end{figure}

The quality of these models can be further evaluated by checking how
well reproduce the first-principles results for the force-constant
bands of strained configurations. As already mentioned,
Fig.~\ref{fig:bands-PTO-strain} shows an essentially perfect agreement
for model $L^{0}$, which is largely preserved in models $L^{I}$ and
$L^{II}$ (the latter is not shown). Naturally, the agreement is worse
for the minimal model $L^{III}$. Figure~\ref{fig:bands-PTO-gs} also
shows the results that model $L^{I}$ gives for the force-constant
bands of PTO's tetragonal ground state, as compared with the
first-principles calculations. As in the fixed-cell cases of
Fig.~\ref{fig:bands-PTO-dist}, it is apparent that the considered
model is not sufficient to render a precise description of all the
bands. Yet, the qualitative agreement is satisfactory.

\subsubsection{Temperature-dependent behavior}

\begin{figure}[t!]
 \includegraphics[width=.9\columnwidth]{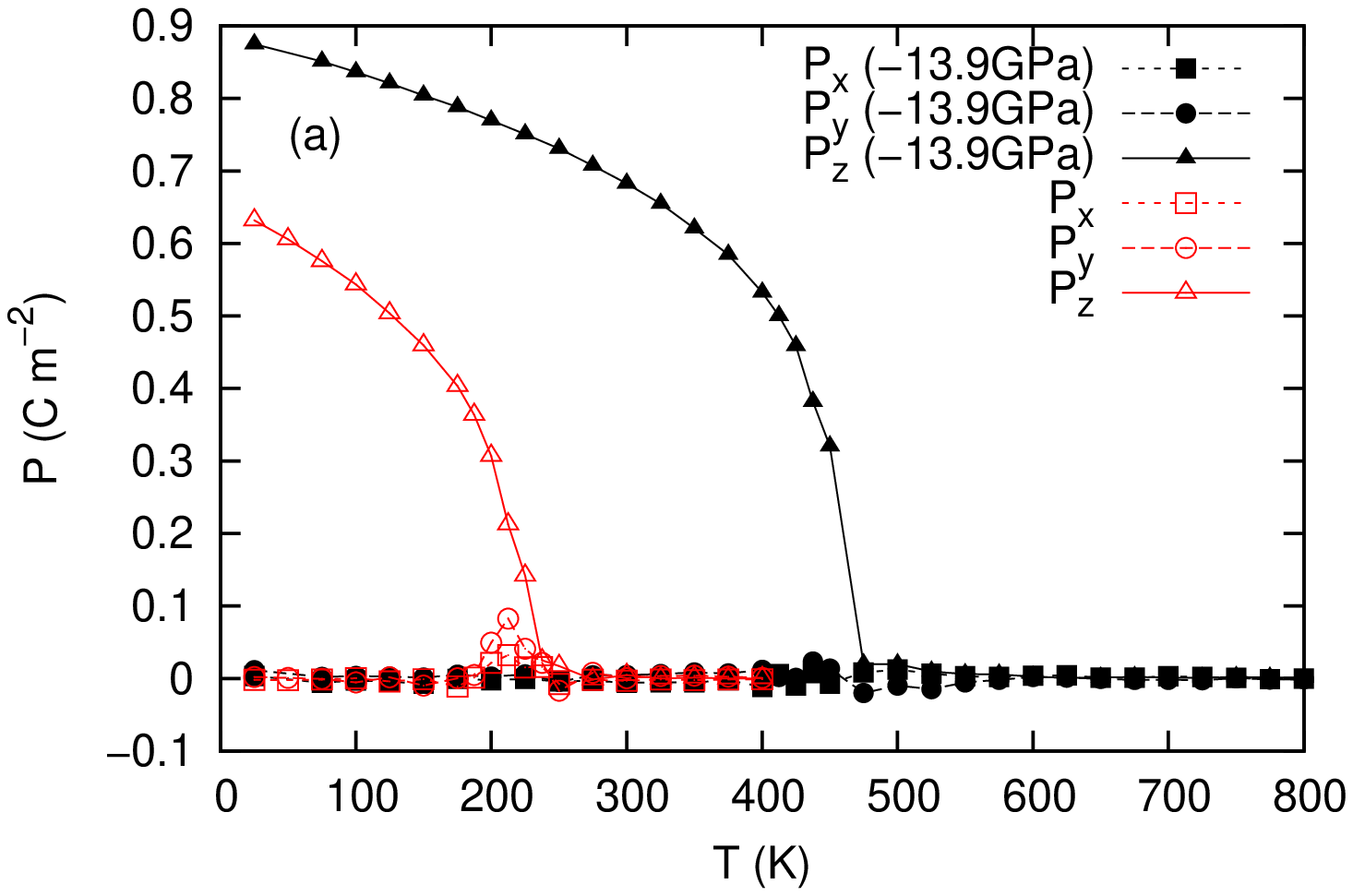} 
 \includegraphics[width=.9\columnwidth]{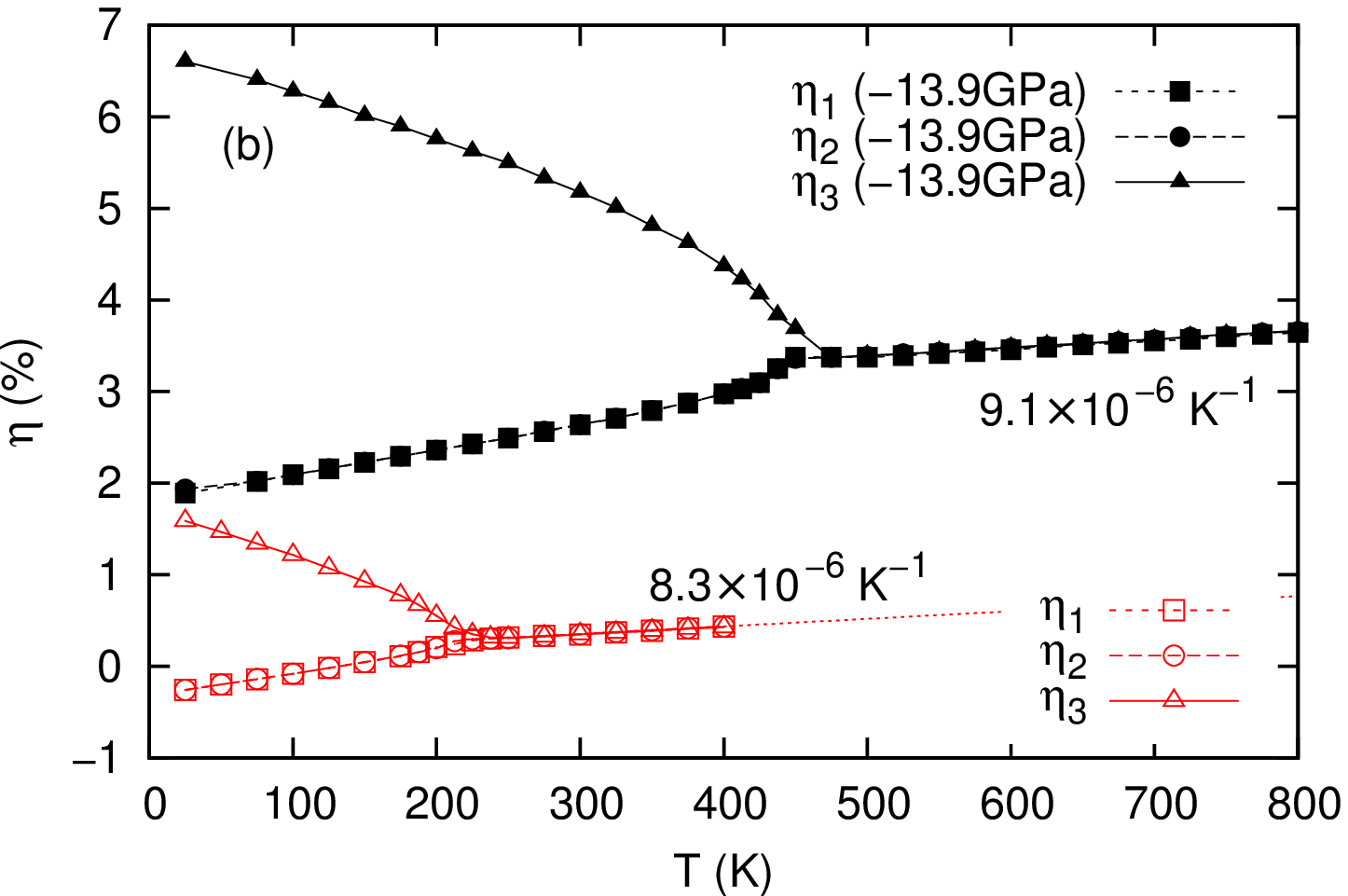}
 \caption{(Color on line.) Temperature-dependent polarization
   [panel~(a)] and strains [panel~(b)] of PTO as calculated from our
   $L^{0}$ potential in two different conditions of external pressure
   (see the text). The LDA-relaxed cubic structure defines the zero of
   strain. The value of the thermal expansion coefficient ($\alpha$)
   of the high-temperature phase is indicated.}
 \label{fig:PTO_tscan_L0}
\end{figure}

We studied the $T$-dependent behavior of our PTO models by running MC
simulations in which both the atomic displacements and strains were
allowed to thermally fluctuate. Figure~\ref{fig:PTO_tscan_L0} shows
the basic results for our $L^{0}$ model when simulated in two
different situations: ($i$) under the condition of zero external
pressure and ($ii$) by imposing an external hydrostatic pressure of
$-$13.9~GPa, which counteracts the underestimation of the LDA result
for the cubic lattice constant. (Taking as a reference the cubic
lattice constant obtained by extrapolating to 0~K the experimental
results in Ref.~\onlinecite{haun87}, this underestimation can be
approximated to be about 2.2\%.) Note that this kind of correction is
customarily made in LDA-based effective-Hamiltonian
works,\cite{zhong94a,zhong95a,waghmare97} and we adopt it here for the
sake of an easier comparison with the literature.

As can be appreciated in Fig.~\ref{fig:PTO_tscan_L0}, our simulated
PTO undergoes a phase transition from the high-$T$ cubic phase to a
low-$T$ tetragonal structure in which one polarization component ($z$
in our default Cartesian setting) becomes different from zero. The
transition is accompanied by a deformation of the cell, which acquires
a $c/a > 1$ aspect ratio. The computed Curie temperature is about
225~K when no external pressure is applied, and increases to about
450~K when we correct for the LDA overbinding. This is the expected
behavior, as it is known that the strength of the FE instabilities in
these perovskite oxides is very sensitive to volume changes (which is
the reason why they have very good piezoelectric properties).

\begin{figure}
 \includegraphics[width=.9\columnwidth]{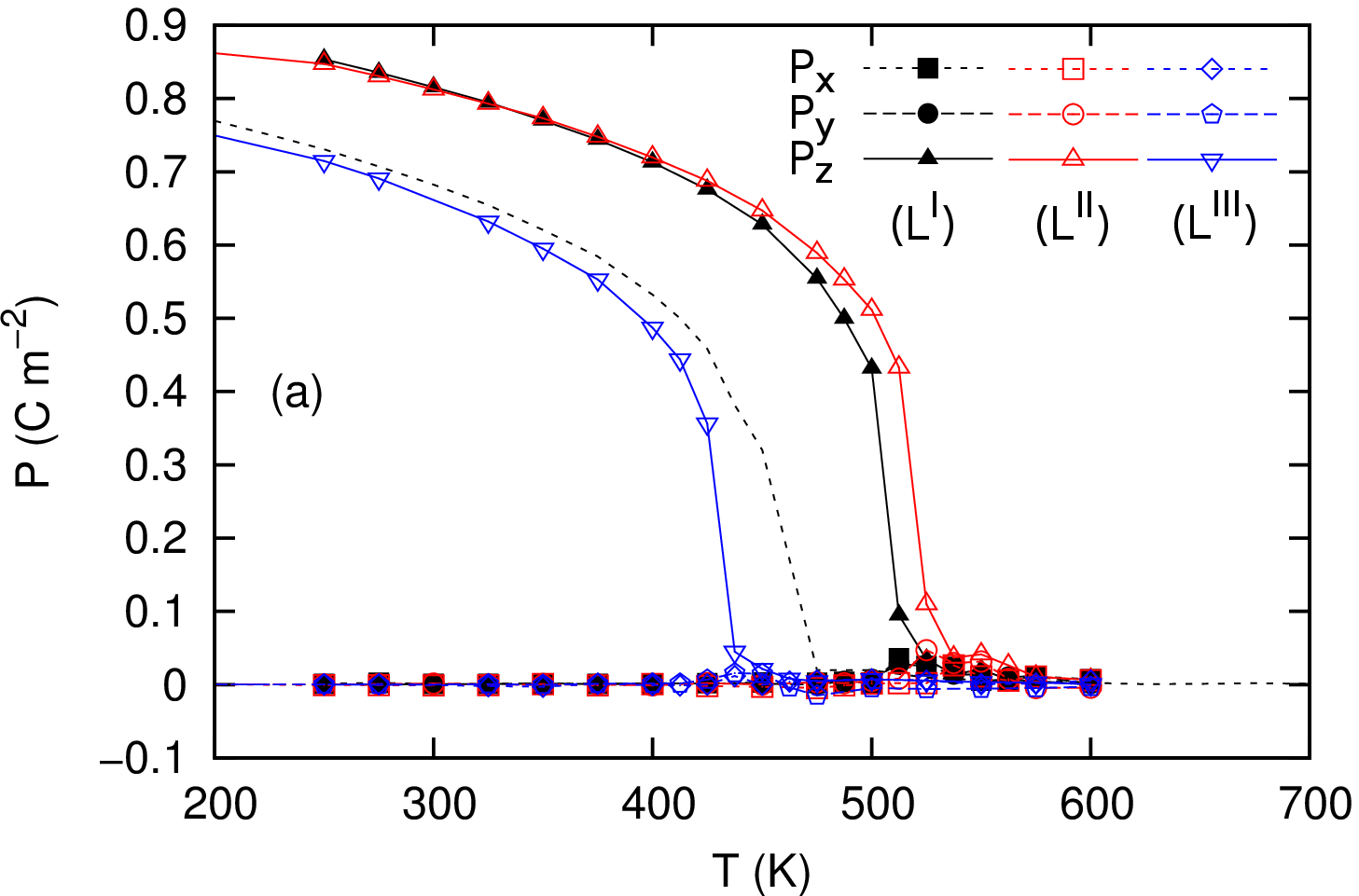} 
 \includegraphics[width=.85\columnwidth]{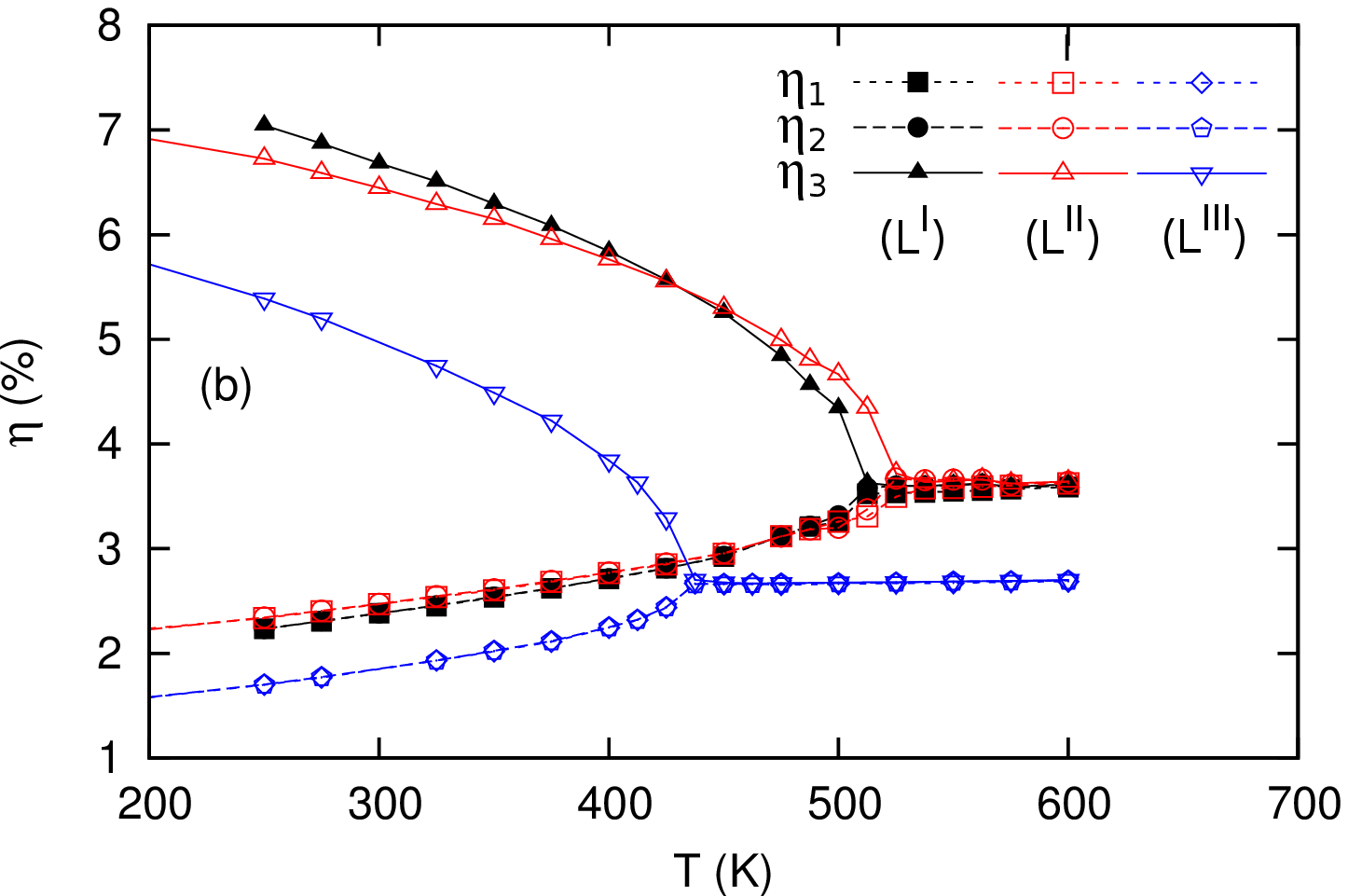} 
 \caption{(Color on line.) Same as
   Fig.~\protect\ref{fig:PTO_tscan_L0}, but for the refined PTO model
   potentials discussed in the text. In all cases an external pressure
   of $-$13.9~GPa is applied. In panel~(a) dashed lines show the
   result for $L^{0}$.}
 \label{fig:PTO_tscan_all}
\end{figure}

Figure~\ref{fig:PTO_tscan_all} shows the results obtained for all the
model potentials listed in Table~\ref{tab:PTO_tetragonal_relaxed}
simulated under the same hydrostatic pressure of
$-$13.9~GPa. Remarkably, in spite of their similarly good description
of the ground state energy and structure, we observe very large
differences in the predicted $T_{\rm C}$'s. It is interesting to note
that, contrary to what we would have
expected,\cite{abrahams68,grinberg04} the obtained $T_{\rm C}$'s do
not correlate well with the energy difference between the ground state
and the RS, nor with the magnitude of the FE distortion. Thus, for
example, the lowest $T_{\rm C}$ (about 440~K) corresponds to the
$L^{III}$ potential, in spite of the fact that the weakest FE
instability ($c/a$~=~1.020; $E_{\rm gs}-E_{\rm RS}$ =
$-$34.5~meV/f.u., where $E_{\rm gs}$ is the ground state energy)
corresponds to the $L^{0}$ model. (The same trends were observed in
the MC runs with no applied pressure.) It is thus clear from these
results that the computed $T_{\rm C}$'s are strongly dependent on
details of the PES that are {\em not} reflected in the energy and
structure of the ground state, a conclusion that can be extended to
all physical properties that we may obtain from our MC
simulations. Hence, the results in Fig.~\ref{fig:PTO_tscan_all}
evidence the critical importance of developing models that include all
the atomic degrees of freedom, and allow for a systematic improvement
of the PES description, if we want to obtain accurate first-principles
results of the thermodynamic properties of materials like PTO.

Let us conclude by giving an additional and striking example of the
importance of {\em hidden} atomistic effects in determining the
macroscopic properties of this material. Our best model for PTO is
probably the one labeled $L^{I}$, which renders a FE transition at
$T_{\rm C}\approx$~510~K. Interestingly, Waghmare and Rabe (WR)
constructed an effective Hamiltonian for PTO, considering only polar
local modes and strains as the model variables, that results in a
significantly higher $T_{\rm C}$ of about 660~K.\cite{waghmare97} At
first sight such a discrepancy may seem surprising, and we made an
effort to understand its origin in some detail. First, we checked that
our model reproduces the energetics of the FE instabilities given by
the WR Hamiltonian rather closely, despite the differences in the
first-principles calculations (e.g., in the pseudopotentials) employed
to compute the parameters. Further, we ran simulations with modified
versions of our model to test subtle features of the WR energy
parametrization (e.g., the inclusion of high-order terms for the polar
local modes), and concluded that they cannot account for the
discrepancy in the computed $T_{\rm C}$.

\begin{figure}
\includegraphics[width=.9\columnwidth]{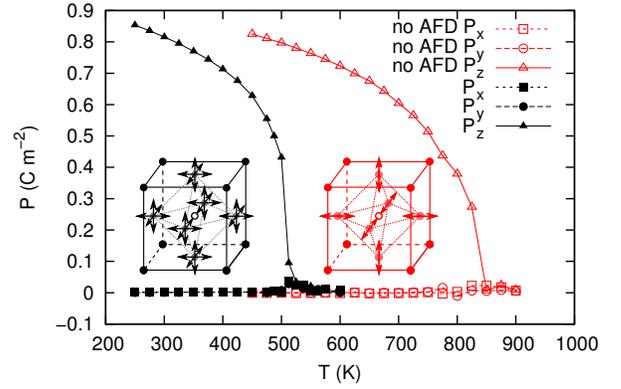}
 \caption{(Color on line.) Temperature-dependent polarization as
   obtained from MC simulations of our $L^{I}$ model. The black solid
   symbols show the results obtained when we allow all possible atomic
   movements (as pictorially depicted for the oxygen atoms in the left
   inset); the red open symbols show the results when we suppress the
   oxygen displacements associated with the rotations of the O$_{6}$
   octahedra (the right inset shows the allowed oxygen displacements
   in this case).}
 \label{fig:t-scan-fixedrotations}
\end{figure} 

We thus turned our attention to the qualitatively distinct features of
our model. Most notably, we describe not only the FE instabilities and
strains, but also the unstable AFD distortions sketched in
Fig.~\ref{fig:instabilities}. It is known that, in most perovskite
oxides, the interaction between FE and AFD modes is a competitive one,
so that they tend to suppress each other.\cite{zhong95b} Hence, to
evaluate the effect of such a competition in our simulated PTO, we ran
simulations in which the O$_{6}$ rotational modes were not allowed. We
imposed this constraint by restricting the motion of the oxygen atoms
as shown in the sketch of Fig.~\ref{fig:t-scan-fixedrotations}. Let us
stress that such a constraint does not affect the energetics
associated with the development of the spontaneous polarization, the
FE ground state being exactly
retained. Figure~\ref{fig:t-scan-fixedrotations} shows the results for
our $L^{I}$ model: In the case without AFDs we got $T_{\rm
  C}\approx$~825~K, which lies about 300~K above the result obtained
from the unconstrained simulation. (The FE-AFD competition was
predicted by other authors to have a similarly large impact on the
$T_{\rm C}$ of the PbZr$_{1-x}$Ti$_{x}$O$_{3}$ solid
solution.\cite{kornev06}) The details of these FE-AFD interactions
will be discussed at length in a future publication. We show this
result here just as a striking example of the physical effects that we
are likely to miss if we restrict ourselves to effective models that,
in spite of looking complete (as e.g. they capture the basic features
of the ground state), may turn out to be too simple.

\subsection{SrTiO$_{3}$}

\subsubsection{Harmonic terms $E_{\rm har}(\{\boldsymbol{u}_{i}\})$ and
  $E_{\rm s}(\boldsymbol{\eta})$}

\begin{figure}[t!]
 \includegraphics[width=\columnwidth]{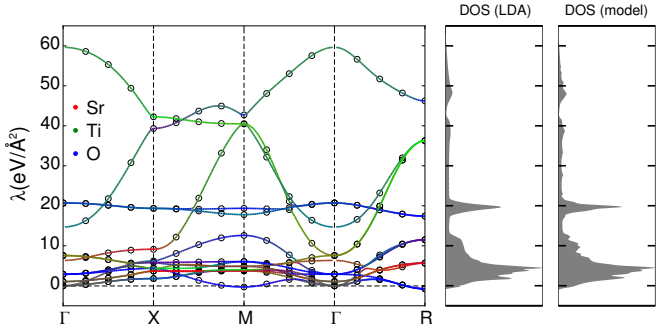}
\caption{(Color on line.) Same as Fig.~\protect\ref{fig:phonons-pto},
  but for SrTiO$_{3}$.}
\label{fig:phonons-sto}
\end{figure}

We extracted all the non-zero harmonic coupling terms from DFPT
calculations\cite{gonze97,wu05} carried out with {\sc
  Abinit}.\cite{abinit09} Representative results for the
force-constant bands of the cubic RS are shown in
Fig.~\ref{fig:phonons-sto}. Note that in this case we only have
AFD-related instabilities; indeed, our LDA calculations render
low-energy, but perfectly stable, FE modes for the cubic phase of STO.

The short-range interatomic interactions that we obtained for STO have
the same spatial extent as those computed for PTO and described
above. As regards the electronic dielectric tensor, Born effective
charges, and harmonic elastic constants, our results for STO are also
analogous to the ones for PTO described above, as the number of
symmetry-independent terms is the same for both materials.

\subsubsection{Fitting $E_{\rm anh}(\{\boldsymbol{u}_{i}\})$}

We fitted the terms in $E_{\rm anh}(\{\boldsymbol{u}_{i}\})$ by
working with a training set of relevant low-symmetry phases that
maintain the cubic STO cell (i.e., with $\boldsymbol{\eta} = 0$). More
precisely, we considered the following AFD-distorted structures:
AFD$_{z}^{i}$ ($P4/mbm$ space group), AFD$_{z}^{a}$ ($I4/mcm$),
AFD$_{xz}^{a}$ ($Imma$, with rotations of equal amplitude about $x$
and $z$) and AFD$_{xyz}^{a}$ ($R\bar{3}c$). As in the case of PTO, we
determined such low-symmetry structures {\sl ab initio} by distorting
the RS according to a specific unstable eigenvector and relaxing the
resulting structure while preserving the targeted symmetry. The
energies and distortion amplitudes computed for these structures are
given in Table~\ref{tab:STO-structures}.

Additionally, in our set we also included two structures generated by
distorting the cubic phase according to the lowest-energy FE
eigendisplacement obtained from our DFPT calculations [which strongly
  resembles the typical FE unstable mode depicted in
  Fig.~\ref{fig:instabilities}(a)]. More precisely, we considered two
distortions involving polarizations along the [001] and [111]
Cartesian directions, respectively. We included such structures in the
training set to capture the anharmonicity of the low-lying FE modes,
which should play a role in determining the non-linear dielectric
response properties of interest in STO.

As in the case of PTO, we worked with a relatively simple model
restricted to pairwise anharmonic interactions extending up to the 4th
order of the Taylor series. Further, we restricted ourselves to
interactions between first-nearest-neighboring atoms, an approximation
that is justified by the rapid spatial decay of the anharmonic
corrections that we observed for STO as well. As we have mentioned
already, these truncations result in the 15 SATs listed in
Table~\ref{tab:SATs}.

\begin{figure}[t!]
 \includegraphics[width=.45\textwidth]{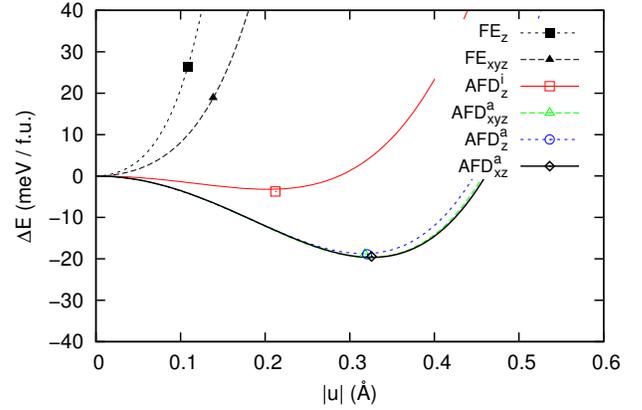} 
 \caption{(Color on line.) Potential-energy wells connecting the RS of
   STO with the low-symmetry phases mentioned in the text. The results
   obtained from our model potential are shown with lines, and the
   points indicate the first-principles results for the energy minima
   or saddles, as well as for the two FE-distorted structures
   considered. All the states shown preserve the cubic cell of the RS
   ($\boldsymbol{\eta} = 0$). The AFD$_{xz}^{a}$ and AFD$_{xyz}^{a}$
   curves are essentially on top of each other and cannot be
   distinguished.}
\label{fig:STO_profile}
\end{figure}

Additionally, in the case of STO we tried to identify the minimal set
of SATs that capture the energetics of the low-symmetry structures in
our training set. We found it possible to do so by considering only
the 4th-order terms with numbers 4, 12, 13, and 15 in
Table~\ref{tab:SATs}. We computed the corresponding parameters by
optimizing $\mathcal{GF}_{E}$ and $\mathcal{GF}_{\nabla E}$, and
obtained the results summarized in Fig.~\ref{fig:STO_profile} and
Table~\ref{tab:STO-structures}. The agreement with the
first-principles data is very good, and we checked that no significant
improvement is obtained by including other SATs listed in
Table~\ref{tab:SATs}.

To construct the present model of STO, we did not take the additional
step of minimizing a goal function $\mathcal{GF}_{\rm hess}$ with
information about the Hessian matrices of the low-energy
structures. It is therefore interesting to check whether the mode
stiffnesses calculated using our effective potential reproduce well
the first-principles data. Representative results are depicted in
Fig.~\ref{fig:DOS-lowsymm}(a), where we show DOS plots constructed
from the force-constant eigenvalues $\lambda_{\boldsymbol{q}s}$ for
the AFD$_{xz}^{a}$ structure, which is the predicted ground state of
the material for $\boldsymbol{\eta} = 0$ (see
Table~\ref{tab:STO-structures}). As it can be seen, our model properly
describes the structure as being a stable one (i.e., we find no modes
with $\lambda_{\boldsymbol{q}s} < 0$) and reproduces well the general
shape of the spectrum. In fact, the overall agreement for the spectrum
of Hessian eigenmodes is comparable to the one obtained for PTO.

\subsubsection{Strain-phonon term $E_{\rm
    sp}(\{\boldsymbol{u}_{i}\},\boldsymbol{\eta})$}

\begin{table*}[t!]    
 \caption{Structural parameters of STO's low-energy phases (see the
   text) as obtained from first-principles LDA calculations and from
   the presented model. A cubic cell with $a$~=~3.845~\AA\ was used in
   all cases, except in the one marked with an asterisk; in that case,
   a full structural relaxation was performed and the resulting
   pseudocubic lattice constants are given. The amplitude of the AFD
   modes is quantified by the corresponding O$_{6}$ rotation angle
   given in degrees (in the AFD$_{xz}^{a}$ and AFD$_{xyz}^{a}$ cases,
   we have equal-magnitude rotations about two and three Cartesian
   axes, respectively; we give the rotation angle for one
   axis.). Energies are given in meV/f.u., taking the result for the
   RS as the zero of energy. Note that the AFD$_{xz}^{a}$ structure
   displays additional small distortions; for example, there are
   anti-polar displacements of the Sr atoms, the off-centering being
   about 0.008~\AA\ and about 0.002~\AA\ for the LDA and model
   calculations, respectively.}
 \begin{tabular}{ccddc}
\hline\hline
 Structure & Method 
           & \multicolumn{1}{c}{Energy} 
           & \multicolumn{1}{c}{O$_{6}$ rot.} & \\
\hline
\multirow{2}{*}{AFD$_{z}^{a}$ ($I4/mcm$)} & LDA    &  -18.9 & 6.7 & \\ 
  & model  &  -18.8 & 6.6 & \\
\multirow{2}{*}{AFD$_{xz}^{a}$ ($Imma$)} & LDA  &  -19.4 & 4.9 & \\ 
       & model &  -20.0 & 4.9 & \\
\multirow{2}{*}{AFD$_{xyz}^{a}$ ($R\bar{3}c$)} & LDA   &  -18.8 & 3.9 & \\
  & model &  -19.7 & 3.9 & \\
\multirow{2}{*}{AFD$_{z}^{i}$ ($P4/mbm$)}  & LDA   &  -3.7 & 4.5 & \\
   & model &  -3.2 & 4.3 & \\
\multirow{2}{*}{AFD$_{z}^{a}$ ($I4/mcm$) *}  &  LDA &    -23.0 & 7.5 &
   $a$~=~$b$~=~3.825~\AA\ , $c$~=~3.869~\AA\ \\
   &  model   &   -23.0 & 7.4 & 
   $a$~=~$b$~=~3.824~\AA\ , $c$~=~3.867~\AA\ \\
\hline
 \end{tabular}
\label{tab:STO-structures}
\end{table*}

\begin{figure}[b!]
\begin{tabular}{cc}
(a) AFD$_{xz}^{a}$ ($\boldsymbol{\eta} = 0$) & (b) AFD$_{z}^{a}$ (relaxed) \\
 \includegraphics[width=.45\columnwidth]{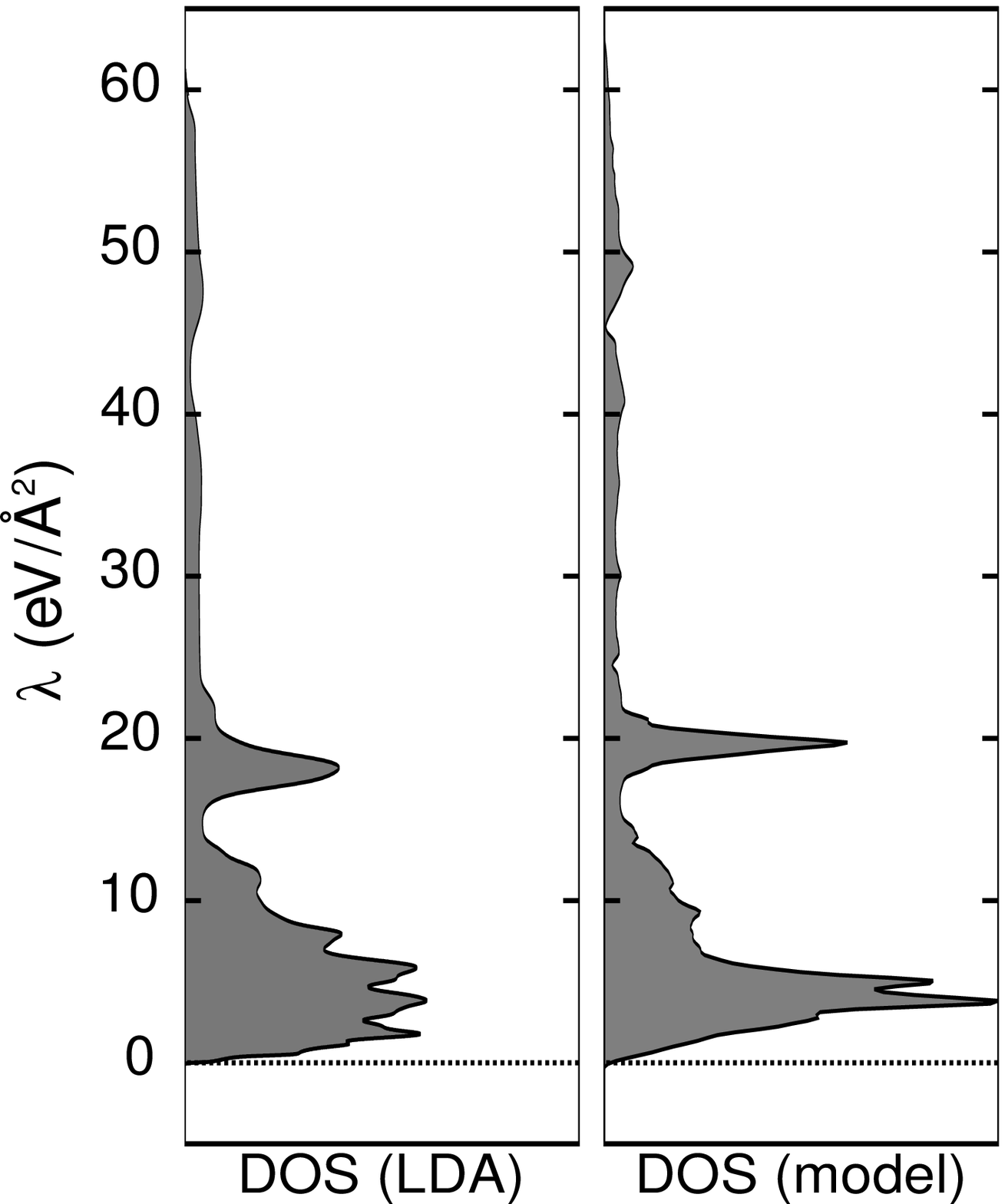} &
\includegraphics[width=.45\columnwidth]{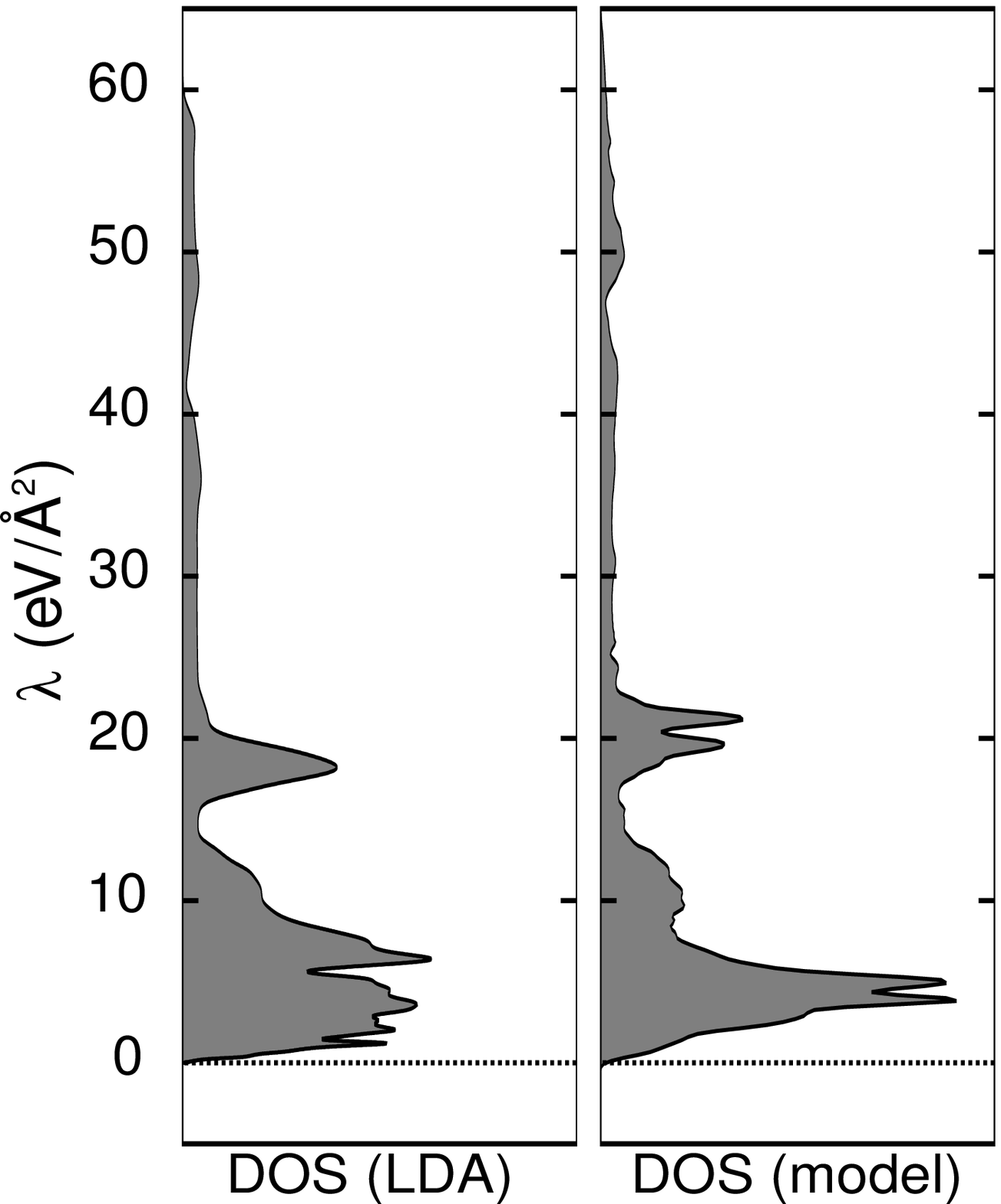} \\
\end{tabular}
 \caption{Density of states (DOS) for two representative low-energy
   phases of STO, obtained in a way that is analogous to the one
   described in the caption of Fig.~\protect\ref{fig:phonons-pto}. We
   show the results obtained from first principles and from the
   presented model potential.}
 \label{fig:DOS-lowsymm}
\end{figure}

As in the case of PTO, we considered only the lowest-order terms that
are allowed by symmetry and capture the most important strain-phonon
effects, which are given by the coefficients
$\boldsymbol{\Lambda}^{(1,2)}$. We computed them directly by employing
the finite-difference approach summarized by
Eqs.~(\ref{eq:finitediff-lambda}) and
(\ref{eq:finitediff-lambda-shortrange}), using strains of $\pm$2\% for
the finite-difference calculations. The resulting model describes the
correct ground state of STO, which is characterized by an
AFD$_{z}^{a}$ distortion ($I4/mcm$ space group). Note that, for
$\boldsymbol{\eta} = 0$, our first-principles calculations indicate
that the ground state is associated with an AFD$_{xz}^{a}$
distortion. Hence, as in PTO's case, the strain-phonon couplings play
a key role in determining the symmetry of the lowest-energy structure;
also like in the case of PTO, such an effect is captured by the
$\boldsymbol{\Lambda}^{(1,2)}$ parameters computed directly {\sl via}
our finite-differences scheme, even though they were not explicitly
fitted to do so.

This model gives an excellent quantitative description of STO's ground
state (see Table~\ref{tab:STO-structures}), and reproduces reasonably
well the corresponding Hessian matrix [see
  Fig.~\ref{fig:DOS-lowsymm}(b)]. Hence, we took this potential as our
effective model for STO, without any further refinement.

\subsubsection{Temperature-dependent behavior}

\begin{figure}
 \includegraphics[width=.45\textwidth]{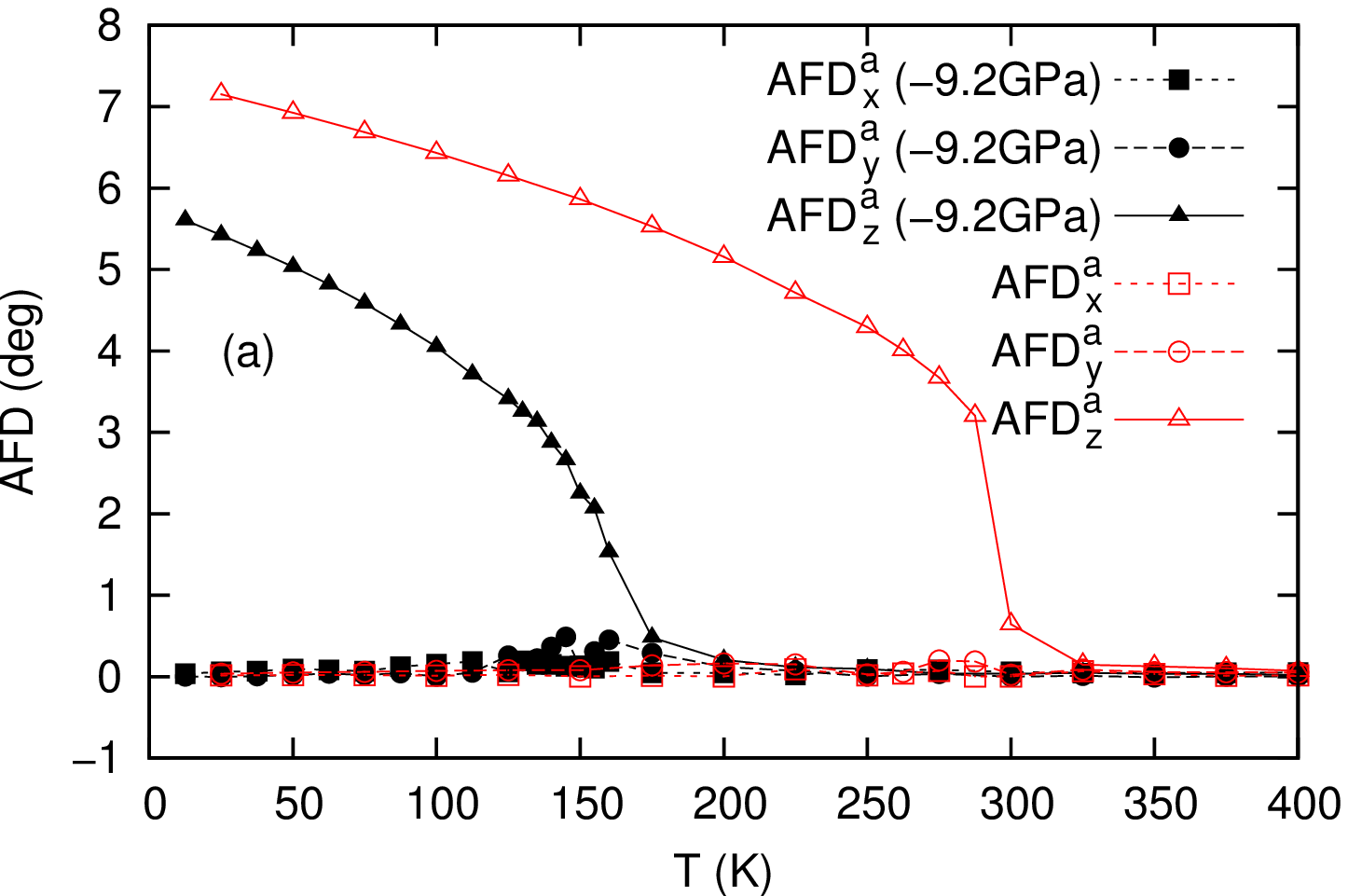} 
 \includegraphics[width=.45\textwidth]{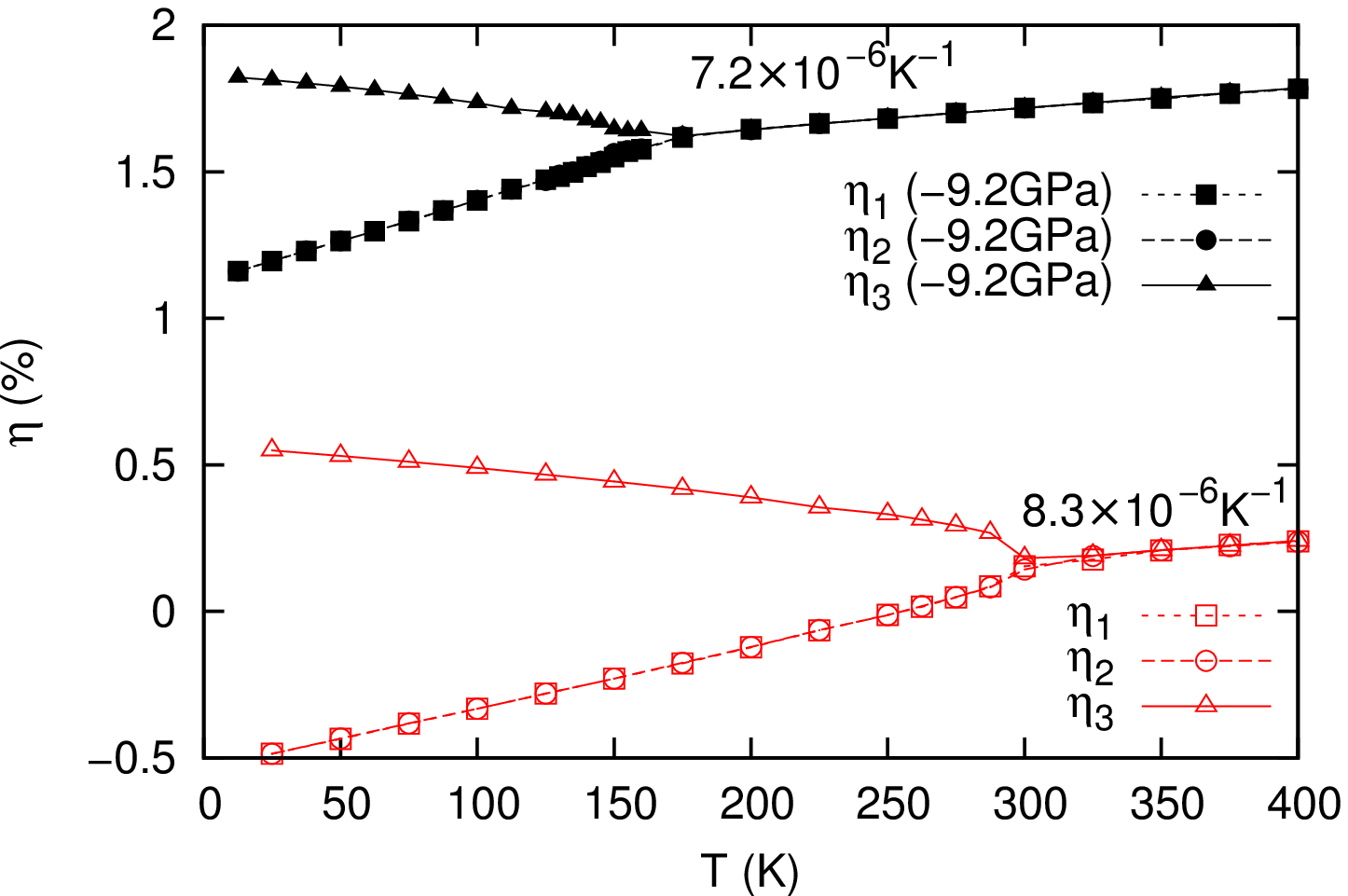} 
 \caption{(Color on line.) Temperature-dependent AFD$^{a}$ distortions
   [panel~(a)] and strains [panel~(b)] of STO as obtained from MC
   simulations of our model. The LDA-relaxed cubic structure defines
   the zero of strain. The value of the thermal expansion coefficient
   ($\alpha$) of the high-temperature phase is indicated in
   panel~(b).}
\label{fig:STO-mcresults}
\end{figure}

Figure~\ref{fig:STO-mcresults} shows the basic results from the MC
simulations performed with our model potential for STO. As in the case
of PTO, we ran simulations ($i$) under zero applied pressure and
($ii$) under an expansive hydrostatic pressure of $-$9.2~GPa, which
approximately corrects for the LDA overbinding. [To compute the
  correction, we used as reference a cubic lattice constant of
  3.90~\AA, obtained by extrapolating to 0~K the experimental results
  for the cubic phase of STO in Ref.~\onlinecite{munakata04}.] In both
cases we get a phase transition from the high-$T$ cubic phase to a
low-$T$ tetragonal structure ($I4/mcm$) with AFD$_{z}^{a}$
character. The transition temperature is about 300~K when no pressure
is applied, and decreases to about 160~K upon application of
$-$9.2~GPa. This is the expected behavior, as the applied pressure is
known to ($i$) reduce the strength of the AFD instabilities and ($ii$)
enhance the FE-AFD competition by softening the FE distortions. These
effects have been studied in previous theoretical works on STO and
related perovskites;\cite{zhong95b,sai00} we have captured them
implicitly (i.e., without any {\sl ad hoc} fitting) when constructing
our model.

\subsection{Discussion}

Let us conclude this Section by commenting on how well our models
reproduce experiment; in particular, let us focus on their performance
to predict one of the most basic properties of these ferroic
materials, namely, the temperature of their structural transition.

The effective-Hamiltonian approach to FE perovskites has been very
successful in reproducing non-trivial behaviors of many complex
materials {\em qualitatively}; examples include the phase diagram of
chemically-disordered solid solutions,\cite{bellaiche00,kornev06} the
occurrence of multiferroic orders,\cite{kornev07,prosandeev13}
strain-\cite{infante10},
finite-size-\cite{ghosez00,naumov04,ponomareva05}, and
electrostatics-\cite{naumov04,ponomareva05} driven effects, and
quantum-phase transitions.\cite{zhong96,iniguez02} However, whenever
the model parameters have been obtained directly from first
principles, and despite the use of pressure corrections as the one
considered here, the quantitative agreement for the predicted
transition temperature has been a poor one. Thus, for example, the
cubic-to-tetragonal transition of BaTiO$_{3}$ was predicted to occur
at about 300~K,\cite{zhong94a,zhong95a} while the experimental result
is 400~K. In the case of KNbO$_{3}$,\cite{krakauer99} the simulations
render a cubic-to-tetragonal transition at 370~K, while the
experimental $T_{\rm C}$ is about 700~K. In this context, the result
of Waghmare and Rabe\cite{waghmare97} for PTO -- i.e., a $T_{\rm C}$
of 660~K that compares reasonably well with the observed value of
760~K -- might be considered as an example of good agreement between
theory and experiment.

The difficulties of the effective-Hamiltonian method to obtain correct
transition temperatures were analyzed in Ref.~\onlinecite{tinte03},
where it was suggested that the discrepancy is to be partly attributed
to an incorrect description of thermal expansion, a problem that is a
direct consequence of the coarse-graining step involved in the
construction of the model. Indeed, the effective Hamiltonians tend to
give an essentially null thermal expansion at high temperatures (see
e.g. Fig.~10 in Ref.~\onlinecite{waghmare97}), which is clearly
against the experimental evidence.

Our models take into account all the degrees of freedom in the
material, and we can thus hope to improve on this aspect. As can be
seen in Fig.~\ref{fig:PTO_tscan_L0}, for PTO we get a thermal
expansion coefficient at high temperatures between
8.3$\times$10$^{-6}$~K$^{-1}$ (from simulations with no applied
pressure) and 9.1$\times$10$^{-6}$~K$^{-1}$ (obtained when pressure is
applied to correct for the LDA overbinding), to be compared with the
experimental result of $12.6\times 10^{-6}$~K$^{-1}$.\cite{haun87}
Similarly, Fig.~\ref{fig:STO-mcresults} shows a thermal expansion
between 7.2$\times$10$^{-6}$~K$^{-1}$ and
8.3$\times$10$^{-6}$~K$^{-1}$ for STO at high temperatures, to be
compared with the value of 8.8$\times$10$^{-6}$~K$^{-1}$ obtained from
experiments.\cite{munakata04} Hence, our models clearly improve the
effective-Hamiltonian description of this effect; yet, the discrepancy
between the computed and measured transition temperatures remains
present.

In the case of PTO, we have found solid evidence that $T_{\rm C}$
depends very significantly on details of the PES that are most often
ignored in theoretical works. In particular, our results strongly
suggest that a realistic model for PTO must necessarily include both
FE and AFD degrees of freedom, as their competition is far from being
negligible. Accordingly, we should probably consider as partly
fortuitous the relatively accurate result obtained for $T_{\rm C}$ in
Ref.~\onlinecite{waghmare97}, where a model without AFDs was employed.

Our results lead to the following important conclusion: We cannot
expect to obtain accurate values for PTO's $T_{\rm C}$ from models
that {\em only} reproduce the basic first-principles results (i.e.,
energy and structure) for the low-symmetry phases of the
material. Further, in order to improve the agreement with experiment,
we should probably extend our model to better reproduce the
lattice-dynamical properties of the key low-energy structures and
other details of the PES. Our results in Fig.~\ref{fig:PTO_tscan_all}
suggest that improvements of that sort, even though they may look like
second-order corrections to the relevant PES, can actually affect the
computed $T_{\rm C}$ by as much as 100~K.

On the other hand, such a strong sensitivity to the details of the
potential has important implications regarding the accuracy required
from the first-principles methods used to compute the model
parameters. Together with the incorrect treatment of thermal
expansion, the authors of Ref.~\onlinecite{tinte03} mentioned DFT
inaccuracy as the second reason to explain the fact that the Curie
temperatures obtained from effective-Hamiltonian simulations are
typically too low as compared with the experimental ones. Their
conjecture was that the FE instabilities obtained from
first principles were too weak, meaning that DFT was probably
underestimating the $|E_{\rm gs}-E_{\rm RS}|$ energy difference
between the RS and the FE ground state. Our results show that, while
important, this is by no means the only characteristic of the PES that
has a large impact on the computed $T_{\rm C}$. Hence, to get accurate
results, we need a first-principles theory that not only describes
correctly the energetics of the FE instability, but also captures
accurately more subtle PES features such as the anharmonic couplings
between different structural distortions, including those that do not
participate directly in the transitions. These are very demanding
requirements for our simulation techniques; thus, it is unclear
whether we presently count with first-principles methods that can
predict an accurate $T_{\rm C}$ for materials like PTO.

Most of the above considerations probably apply to STO as well. Yet,
as the transition occurs at a relatively low temperature in this case,
an additional factor must be taken into account. As demonstrated in a
variety of theoretical works,\cite{zhong96,iniguez02,akbarzadeh04} in
order to get a precise calculation of the structural transition
temperatures of ABO$_3$ perovskites, it is important to consider the
quantum (i.e., wave-like) character of the atoms. Indeed, Zhong and
Vanderbilt simulated STO at both the classical and quantum-mechanical
levels, using an effective Hamiltonian constructed from first
principles, and found that quantum fluctuations shift down the
cubic-to-tetragonal transition by about 20~K.\cite{zhong96} (Quantum
effects will typically promote disorder, and thus result in reduced
transition temperatures.) Hence, in the case of STO, we can assume
that part of the discrepancy between our computed transition
temperature ($\sim$~160~K) and the experimental one (105~K) comes from
the fact we treated atoms as classical objects in our MC
simulations. Finally, let us note that our computed transition
temperature seems consistent with the result of about 130~K reported
in Ref.~\onlinecite{zhong96} for the classical case.

\section{Summary and conclusions}

We have proposed a new method for the construction of first-principles
model potentials that permit large-scale simulations of
lattice-dynamical phenomena. Our scheme mimics the traditional
approach to lattice dynamics in solid-state textbooks, i.e., we start
from a suitably chosen reference structure (RS) and express the energy
of the material as a Taylor series for the structural distortions of
such a RS. There are many advantages in adopting such a simple
approach; most notably, our potentials can be trivially formulated for
any compound, and their ability to reproduce the first-principles data
can be improved in a systematic and well-defined way. Further, most of
the potential parameters correspond to the usual (elastic,
force-constant, etc.) tensors discussed in condensed-matter theory,
which allows for a transparent physical interpretation.

We have described the details of such an approach, and proposed a
practical strategy to compute the model parameters from first
principles. Our method is especially convenient in that regard too, as
we have shown that many of the key model parameters (e.g., all the
couplings at the harmonic level) can be readily obtained from
density-functional-perturbation-theory calculations that are widely
available today.

We have illustrated our method with applications to two especially
challenging cases, namely, ferroic oxides PbTiO$_{3}$ and
SrTiO$_{3}$. These materials undergo structural phase transitions
driven by soft phonon modes, which implies that the potential-energy
surface (PES) that our models have to capture is strongly
anharmonic. We have discussed in detail the case of PTO, where the
large structural deformations involved in the ferroelectric phase
transition make it especially challenging to construct a
quantitatively accurate model. Moreover, we have solved our PTO
potential by means of Monte Carlo simulations and discovered a variety
of unexpected effects, ranging from novel structural phases when the
strain deformations are constrained to a surprisingly strong
dependence of the computed Curie temperature on the details of the
PES. The case of STO turned out to be much easier to tackle and led to
quantitatively more accurate predictions, probably because the
structural distortions involved in its ferroic transformation are
smaller. The connections of our method with the so-called {\sl
  first-principles effective Hamiltonian} approach to the study of
temperature-driven effects in ferroic perovskite oxides -- which was
introduced about 20 years ago, and of which our scheme can be
considered a natural extension and generalization -- have been
discussed in some detail.

We believe that our effective potentials can be used to great
advantage in the investigation of the thermodynamic properties of
(meta)stable material phases, which are largely dominated by harmonic
effects that our models describe with first-principles accuracy. While
we have not considered any such case in this work, we believe that the
demonstrated ability of our models to deal with strongly anharmonic
effects suggests that their application to (the much simpler)
quasi-harmonic cases will be a very successful one. Hence, we hope the
current methodology can become a standard tool for large-scale
simulations of the lattice-dynamical properties of materials at
realistic operating conditions of temperature, pressure, etc.

Work supported by the EC-FP7 project OxIDes (Grant No. CP-FP
228989-2), MINECO-Spain (Grants No. MAT2010-18113,
No. MAT2010-10093-E, and No. CSD2007-00041) and CSIC [JAE-doc program
  (JCW)]. We made use of the facilities provided by the CESGA
Supercomputing Center. Discussions with L. Bellaiche,
C. Escorihuela-Sayalero, A.~Garc\'{\i}a, P.~Garc\'{\i}a-Fern\'andez,
and M.~Stengel are gratefully acknowledged.

\bibliography{biblio}

\begin{thebibliography}{63}
\bibitem [{fns()}]{fnsilicon}%
  \BibitemOpen
  \href@noop {} {}\bibinfo {note} {In the case of silicon and other monoatomic
  compounds, all the Born effective charges are identically zero by symmetry.
  In such cases, even if the material is insulating, we do not have harmonic
  dipole-dipole interactions.}\BibitemShut {Stop}%
\end{thebibliography}

\end{document}